\documentclass[draft,tightenlines,nofootinbib,preprint,aps,eqsecnum,amsmath,amssymb]{revtex4}

\newcommand{\beq}{\begin{equation}}
\newcommand{\eeq}{\end{equation}}
\newcommand{\bea}{\begin{eqnarray}}
\newcommand{\eea}{\end{eqnarray}}

\newcommand{\sgn}{\mbox{\boldmath $\epsilon$}}

\begin{document}

\title{The Chrono-geometrical Structure of Special and General
Relativity: a Re-Visitation of Canonical Geometrodynamics.}

\medskip

\author{Luca Lusanna}

\affiliation{ Sezione INFN di Firenze\\ Polo Scientifico\\ Via Sansone 1\\
50019 Sesto Fiorentino (FI), Italy\\ Phone: 0039-055-4572334\\
FAX: 0039-055-4572364\\ E-mail: lusanna@fi.infn.it}

\bigskip
\bigskip

\begin{abstract}

A modern re-visitation of the consequences of the lack of an
intrinsic notion of instantaneous 3-space in relativistic theories
leads to a reformulation of their kinematical basis emphasizing
the role of non-inertial frames centered on an arbitrary
accelerated observer. In special relativity the exigence of
predictability implies the adoption of the 3+1 point of view,
which leads to a well posed initial value problem for field
equations in a framework where the change of the convention of
synchronization of distant clocks is realized by means of a gauge
transformation. This point of view is also at the heart of the
canonical approach to metric and tetrad gravity in globally
hyperbolic asymptotically flat space-times, where the use of
Shanmugadhasan canonical transformations allows the separation of
the physical degrees of freedom of the gravitational field (the
tidal effects) from the arbitrary gauge variables. Since a global
vision of the equivalence principle implies that only global
non-inertial frames can exist in general relativity, the gauge
variables are naturally interpreted as generalized relativistic
inertial effects, which have to be fixed to get a deterministic
evolution in a given non-inertial frame. As a consequence, in each
Einstein's space-time in this class the whole chrono-geometrical
structure, including also the clock synchronization convention, is
dynamically determined and a new approach to the Hole Argument
leads to the conclusion that "gravitational field" and
"space-time" are two faces of the same entity. This view allows to
get a classical scenario for the unification of the four
interactions in a scheme suited to the description of the solar
system or our galaxy with a deperametrization to special
relativity and the subsequent possibility to take the
non-relativistic limit.

\bigskip
Lectures given at the 42nd Karpacz Winter School of Theoretical
Physics, "Current Mathematical Topics in Gravitation and
Cosmology", Ladek, Poland, 6-11 February 2006.

\end{abstract}

\maketitle

\vfill\eject

\section{Introduction}

The main theoretical challenge in the next future will be to find
a classical scenario allowing a unified description of gravity
with the standard model of elementary particles (or its
extensions), so to clarify the problems to be faced in trying to
reconcile general relativity and quantum theory.
\medskip

On one side particle physics is described by gauge theories in the
inertial frames of Minkowski space-time. This implies the
following conceptual problems:\medskip

\noindent A) The need of Einstein's convention for the
synchronization of distant clocks (i.e. the definition of a notion
of instantaneous 3-space, which is not intrinsically given in
special relativity differently from the conformal structure
identifying the allowed paths of light rays) to be made by the
inertial observer describing the phenomena consistently with the
relativity principle and the light postulates. However, all
realistic observers are accelerated and there is no consensus on
how to define relativistic non-inertial frames and on how
non-inertial observers describe the dynamics. This fact, together
with the need of a well-posed Cauchy problem for classical field
equations like Maxwell's ones to have predictability, forces the
adoption of the 3+1 point of view, of more general clock
synchronization conventions and of a formulation in which the
change of such conventions is realized by means of a gauge
transformation not changing the physics but only its description
(the appearances of phenomena).\hfill\break

\noindent B) The need of an implementation of the Poincare' group
on either the configuration or the phase space of the relativistic
system, with all the complications implied by the Lorentz
signature of Minkowski space-time. As a consequence, relativistic
kinematics is highly non trivial (think to the problem of the
relativistic center of  mass), so that relativistic particle
mechanics, extended objects, fluids, relativistic statistical
mechanics, classical field theory turn out to be extremely more
complicated than their non-relativistic counterparts.\hfill\break

\noindent C) The gauge principle (the minimal coupling) together
with manifest Lorentz covariance force us to introduce redundant
non-physical gauge variables with the associated invariance under
local Lie groups of gauge transformations acting on some internal
space. As a consequence, all relativistic Lagrangians are
singular, the action principles have to be studied with the second
Noether theorem, the Hamiltonian formulation requires Dirac's
theory of constraints and the measurable quantities are the gauge
invariant Dirac observables (DO).

\bigskip

On the other hand, the relativistic description of gravity given
by general relativity abandons the relativity principle and
replaces it with the equivalence principle. Special relativity can
be recovered only locally by a freely falling observer in a
neighborhood where tidal effects are negligible. As a consequence,
global inertial frames are not admitted and Einstein's geometrical
view of the gravitational field implies the the 4-metric of
space-time, the dynamical field mediating the gravitational
interaction, also determines the line element (i.e. the
chrono-geometrical structure) of the space-time. Therefore, the
4-metric teaches relativistic causality to all the other fields:
now the conformal structure (the allowed  paths of light rays) is
point-dependent. The geometrical view is intimately connected with
the principle of general covariance: the form-invariance of
Einstein's equations in every 4-coordinate system (invariance
under active diffeomorphisms) requires that all the fields have a
tensorial character. But this implies the invariance of the action
principle under passive diffeomorphisms (ordinary coordinate
transformations). The shift from the Hilbert action to the ADM
one, needed for the Hamiltonian formulation, replaces the passive
diffeomorphisms with local Noether transformations and again the
second Noether theorem implies the need of Dirac's theory of
constraints in phase space. The canonical formalism and the
absence of global inertial frames require the 3+1 point of view as
in special relativity: as a consequence the allowed space-times
must be globally hyperbolic and the gauge equivalence of the
allowed non-inertial frames is now implied  by general covariance.
The superiority of the manifestly covariant configuration approach
(no choice of what is {\it time} in a relativistic framework) is
illusory: to have predictability we are forced to face the Cauchy
problem of Einstein's equations and only the Hamiltonian approach
has the tools to do it. Moreover, the space-time must be
asymptotically flat and without super-translations, so that the
asymptotic ADM Poincare' generators are well defined: when the
Newton constant is switched off they must tend to the Poincare'
generators of special relativity so that ordinary particle physics
can be recovered. In this class of (singularity-free) space-times
there are asymptotic inertial observers to be identified with the
fixed stars and there is a real temporal evolution (absence of the
{\it frozen picture}) governed by the ADM energy. The ADM energy
density is coordinate-dependent (the energy problem in general
relativity), because it depends on the 8 gauge variables hidden in
the 4-metric and its time-derivative. However, in the framework of
Dirac's theory of constraints, the Shanmugadhasan canonical
transformations allow to separate the two pairs of canonical
variables describing the {\it tidal effects} from the gauge
variables associated with the {\it relativistic generalized
inertial effects} off-shell, i.e. {\it before} solving Hamilton
equations. Only after a complete gauge fixing, namely after the
choice of a  non-inertial frame with its pattern of relativistic
inertial forces, the Hamilton equations for the tidal degrees of
freedom and matter (if present) become deterministic. Their
solution with suitable Cauchy data identifies a non-inertial
4-coordinate system for the Einstein space-time, whose
chrono-geometrical structure (including the clock synchronization
convention) is dynamically determined.
\bigskip

These lectures are an introduction to a modern treatment of these
topics starting from the Galilei space-time (Section II) and then
discussing Minkowski space-time (Section III) and the previous
class of Einstein space-times (Section IV).

In the Appendix there are some notions on Dirac's theory of
constraints.

\medskip

While in Ref.\cite{1a} there is a review of constraint theory
(including the Shanmugadhasan canonical transformation and the
notion of Dirac observable) and of its applications to gauge
models in Minkowski space-time, in Refs. \cite{2a,3a} there is a
detailed description of general relativity from this canonical
point of view.

\section{The Chrono-Geometrical Structure of Newton Physics:
Galilei Space-Time.}

In Newton physics there are separated absolute notions of {\it
time} and {\it space}, so that we can speak of absolute
simultaneity and of instantaneous Euclidean 3-spaces with the
associated Euclidean spatial distance notion. This non-dynamical
chrono-geometrical structure is formalized in the so called
Galilei space-time. The Galilei relativity principle assumes the
existence of preferred inertial frames with inertial Cartesian
coordinates, where free bodies move along straight lines (Newton's
first law) and Newton's equations take the simplest form. In
Galilei space-time inertial frames centered on inertial observers
are connected by the kinematical group of Galilei transformations.
In Newton gravity the equivalence  principle states the equality
of inertial and gravitational mass. In non-inertial frames
inertial (or fictitious) forces proportional to the mass of the
body appear in Newton's equations.
\bigskip

For isolated systems the 10 generators of the Galilei group are
Noether constants of motion. The Abelian nature of the Noether
constants (the 3-momentum) associated to the invariance under
translations allow to make a global separation of the center of
mass from the relative variables (usually the Jacobi coordinates,
identified by the centers of mass of subsystems, are preferred):
in phase space this can be done with canonical transformations {\
point} both in the coordinates and in the momenta. Also the
conserved Galilei boosts identify the center of mass. Instead the
non-Abelian nature of the Noether constants (the angular momentum)
associated with the invariance under rotations implies that there
is no unique separation \cite{4a} of the relative variables in 6
orientational ones (the body frame in the case of rigid bodies)
and in the remaining vibrational (or {\it shape}) ones. As a
consequence, an isolated deformable body or a system of particles
may rotate by changing the shape (the falling cat, the diver). In
Ref.\cite{5a} there is a treatment of this part of the kinematics
by means of canonical spin bases and of dynamical body frames,
which can be extended to the relativistic case where the notions
of Jacobi coordinates, reduced masses and tensors of inertia are
absent and can be recovered only when extended bodies are
simulated with multipolar expansions.

\bigskip

Other non-conventional aspects of non-relativistic physics are:
\medskip
\noindent A) The many-time formulation of classical particle
dynamics \cite{6a} with as many first class constraints as
particles. Like in the special relativistic case a distinction
arises between physical positions and canonical configuration
variables and a non-relativistic version of the no-interaction
theorem emerges.\hfill\break

\noindent B) The possibility to gauge the kinematical and internal
symmetry groups of an extended system (the simplest one being the
Galilean one-time and two-time harmonic oscillator) at the
position of the center of mass to define couplings to external
gauge fields \cite{7a}.\hfill\break

\noindent C) The possibility to define standard and generalized
Newtonian gravity theories as gauge theories of the extended
Galilei group \cite{8a} by studying the non-relativistic limit of
the ADM action of metric gravity.\hfill\break

\noindent D) The quantum mechanics of particles in non-rigid
non-inertial frames \cite{9a}.

\section{The Chrono-Geometrical Structure of Special Relativity: Minkowski
Space-Time, Clock Synchronization, Non-Inertial Frames,
Parametrized Minkowski Theories.}

As a consequence of Einstein's Annus Mirabilis 1905, all special
relativistic physical systems, defined in the inertial frames of
Minkowski space-time, are manifestly covariant under the
transformations of the kinematical Poincare' group connecting
inertial frames ({\it relativity principle}). Minkowski space-time
has an absolute (namely non-dynamical) chrono-geometrical
structure. The {\it light postulates} say that the two-way (or
round trip) velocity of light (only one clock is needed for its
definition) is $c$, namely it is i) constant and ii) isotropic.
The Lorentz signature of its 4-metric tensor implies that every
time-like observer can identify the light-cone (the conformal
structure, i.e. the locus of the trajectories of light rays) in
each point of the world-line. But there is {\it no notion of an
instantaneous 3-space, of a spatial distance and of a one-way
velocity of light between two observers} (the problem of the
synchronization of distant clocks). Since the relativity principle
privileges inertial observers and Cartesian coordinates $x^{\mu} =
(x^o = c t; \vec x)$ with the time axis centered on them (inertial
frames), the $x^o = const.$ hyper-planes of inertial frames are
usually taken as Euclidean instantaneous 3-spaces, on which all
the clocks are synchronized. Indeed they can be selected with {\it
Einstein's convention} for the synchronization of distant clocks
to the clock of an inertial observer. This inertial observer $A$
sends a ray of light at $x^o_i$ to a second accelerated observer
B, who reflects it towards A. The reflected ray is reabsorbed by
the inertial observer at $x^o_f$. The convention states that the
clock of B at the reflection point must be synchronized with the
clock of A when it signs ${1\over 2}\, (x^o_i + x^o_f)$. This
convention selects the $x^o = const.$ hyper-planes of inertial
frames as simultaneity 3-spaces and implies that with this
synchronization the two-way (A-B-A)  and one-way (A-B or B-A)
velocities of light coincide and the spatial distance between two
simultaneous point is the (3-geodesic) Euclidean distance.

\medskip

However, real observers are never inertial and for them Einstein's
convention for the synchronization of clocks is not able to
identify globally defined simultaneity 3-surfaces, which could
also be used as Cauchy surfaces for Maxwell equations. The 1+3
{\it point of view} tries to solve this problem starting from the
local properties of an accelerated observer, whose world-line is
assumed to be the time axis of some frame. Since only the observer
4-velocity is given, this only allows to identify the tangent
plane of the vectors orthogonal to this 4-velocity in each point
of the world-line. Then, both in special and general relativity,
this tangent plane is identified with an instantaneous 3-space and
3-geodesic Fermi coordinates are defined on it and used to define
a notion of spatial distance. However this construction leads to
{\it coordinate singularities}, because the tangent planes in
different points of the world-line will intersect each other at
distances from the world-line of the order of the (linear and
rotational) {\it acceleration radii} of the  observer. Another
type of coordinate singularity arises in all the proposed {\it
uniformly rotating} coordinate systems: if $\omega$ is the
constant angular velocity, then at a distance $r$ from the
rotation axis such that $\omega\, r = c$, the ${}^4g_{oo}$
component of the induced 4-metric vanishes. This is the so-called
{\it horizon problem for the rotating disk}: the time-like
4-velocity of an observer sitting on a point of the disk becomes
light-like in this coordinate system when $\omega\, r = c$.

While in particle mechanics one can formulate a theory of
measurement for accelerated observers based on the {\it locality
hypothesis} \footnote{Standard clocks and rods do not feel
acceleration and at each instant the detectors of the
instantaneously comoving inertial observer give the correct
data.}, this methodology does not work with moving continuous
media (for instance the constitutive equations of the
electromagnetic field inside them in non-inertial frames are still
unknown) and in presence of electromagnetic fields when their
wavelength is comparable with the acceleration radii of the
observer (the observer is not enough "static" to be able to
measure the frequency of such a wave).
\medskip

See Refs. \cite{2a,10a} for a review of these topics.

\bigskip

This state of affairs and the  need of predictability (a
well-posed Cauchy problem for field theory) lead to the necessity
of abandoning the 1+3 point of view and to shift to the 3+1 one.
In this point of view, {\it besides the world-line of an arbitrary
time-like observer, it is given a global 3+1 splitting of
Minkowski space-time}, namely a foliation of it whose leaves are
space-like hyper-surfaces. Each leaf is {\it both} a Cauchy
surface for the description of physical systems and an
instantaneous (in general Riemannian) 3-space, namely a notion of
simultaneity implied by a clock synchronization convention
different from Einstein's one.

Even if it is unphysical to give initial data on a non-compact
space-like hyper-surface, this is the only way to be able to use
the existence and uniqueness theorem for the solutions of partial
differential equations. In the more realistic mixed problem, in
which we give initial data on the Earth and we add an arbitrary
information on the null boundary of the future causal domain of
the Earth (that is we prescribe the data arriving from the rest of
the universe, the ones observed by astronomers), the theorem
cannot be shown to hold!

\bigskip

The extra structure of the 3+1 splitting of Minkowski space-time
allows to enlarge its atlas of 4-coordinate systems with the
definition of {\it Lorentz-scalar observer-dependent radar
4-coordinates} $\sigma^A = (\tau ; \sigma^r)$, $A = \tau , r$.
Here $\tau$ is either the proper time of the accelerated observer
or any monotonically increasing function of it, and is used to
label the simultaneity leaves $\Sigma_{\tau}$ of the foliation. On
each leaf $\Sigma_{\tau}$ the point of intersection with the
world-line of the accelerated observer is taken as the origin of
curvilinear 3-coordinates $\sigma^r$, which can be assumed to be
globally defined since each $\Sigma_{\tau}$ is diffeomorphic to
$R^3$. To the coordinate transformation $x^{\mu} \mapsto \sigma^A$
($x^{\mu}$ are the standard Cartesian coordinates) is associated
an inverse transformation $\sigma^A \mapsto x^{\mu} = z^{\mu}(\tau
, \sigma^r)$, where the functions $z^{\mu}(\tau , \sigma^r)$
describe the embedding of the simultaneity surfaces
$\Sigma_{\tau}$ into Minkowski space-time. The 3+1 splitting leads
to the following induced 4-metric (a functional of the embedding):
${}^4g_{AB}(\tau , \sigma^r) = {{\partial z^{\mu}(\sigma )}\over
{\partial \sigma^A}}\, {}^4\eta_{\mu\nu}\, {{\partial
z^{\nu}(\sigma )}\over {\partial \sigma^B}} = {}^4g_{AB}[z(\sigma
)]$, where ${}^4\eta_{\mu\nu} = \epsilon\, (+ - - -)$ with
$\epsilon = \pm 1$ according to particle physics or general
relativity convention respectively. The quantities
$z^{\mu}_A(\sigma ) = {{\partial z^{\mu}(\sigma )}\over {\partial
\sigma^A}}$ are cotetrad fields on Minkowski space-time.

\bigskip

To avoid the previously quoted coordinate singularities, an
admissible 3+1 splitting of Minkowski space-time must have the
embeddings $z^{\mu}(\tau ,\sigma^r)$ of the space-like leaves
$\Sigma_{\tau}$ of the associated foliation satisfying the
M$\o$ller conditions on the coordinate transformation \cite{11a}

\begin{eqnarray*}
 && \epsilon\, {}^4g_{\tau\tau}(\sigma ) > 0,\qquad
 \epsilon\, {}^4g_{rr}(\sigma ) < 0,\qquad \begin{array}{|ll|} {}^4g_{rr}(\sigma )
 & {}^4g_{rs}(\sigma ) \\ {}^4g_{sr}(\sigma ) &
 {}^4g_{ss}(\sigma ) \end{array}\, > 0, \qquad
 \epsilon\, det\, [{}^4g_{rs}(\sigma )]\, < 0,\nonumber \\
 &&{}\nonumber \\
 &&\Rightarrow det\, [{}^4g_{AB}(\sigma )]\, < 0.
 \nonumber \\
 \end{eqnarray*}

Moreover, the requirement that the foliation be well defined at
spatial infinity may be satisfied by asking that each simultaneity
surface $\Sigma_{\tau}$ tends to a space-like hyper-plane there,
namely we must have $z^{\mu}(\tau ,\sigma^r)\, \rightarrow\,
x^{\mu}(0) + \epsilon^{\mu}_A\, \sigma^A$ for some set of
orthonormal asymptotic tetrads $\epsilon^{\mu}_A$.

\bigskip

As a consequence, any admissible 3+1 splitting leads to the
definition of a {\it non-inertial frame centered on the given
time-like observer} and coordinatized with Lorentz-scalar
observer-dependent radar 4-coordinates. While  inertial frames
centered on inertial observers are connected by the
transformations of the Poincare' group, the non-inertial ones are
connected by {\it passive frame-preserving diffeomorphism}: $\tau
\mapsto \tau^{'}(\tau ,\sigma^r)$, $\sigma^r \mapsto \sigma^{{'}\,
r}(\sigma^s)$. It turns out that M$\o$ller conditions forbid
uniformly rotating non-inertial frames: {\it only differentially
rotating ones are allowed} (the ones  used by astrophysicists in
the modern description of rotating stars). In  Refs.\cite{10a}
there is a detailed discussion of this topic and there is the
simplest example of  3+1 splittings whose leaves are space-like
hyper-planes carrying admissible differentially rotating
3-coordinates ($\sigma = |\vec \sigma |$; $\epsilon^{\mu}_r$ are
asymptotic space-like axes; $\alpha_i$, $i=1,2,3$, are Euler
angles):

\begin{eqnarray*}
 z^{\mu}(\tau ,\vec \sigma ) &=& x^{\mu}(\tau ) + \epsilon^{\mu}_r\,
R^r{}_s(\tau , \sigma )\, \sigma^s,\nonumber \\
 &&{}\nonumber \\
 &&R^r{}_s(\tau ,\sigma ) {\rightarrow}_{\sigma \rightarrow
 \infty} \delta^r_s,\qquad \partial_A\, R^r{}_s(\tau
 ,\sigma )\, {\rightarrow}_{\sigma \rightarrow
 \infty}\, 0,\nonumber \\
 &&{}\nonumber \\
 R^r{}_s(\tau ,\sigma ) &=& R^r{}_s(\alpha_i(\tau,\sigma )) =
 R^r{}_s(F(\sigma )\, {\tilde \alpha}_i(\tau)),\nonumber \\
 &&{}\nonumber \\
 &&0 < F(\sigma ) < {1\over {A\, \sigma}},\qquad {{d\, F(\sigma
 )}\over {d\sigma}} \not= 0.
 \end{eqnarray*}

\noindent Each function $F(\sigma )$, satisfying the  M$\o$ller\,
conditions given in the last line, defines an admissible
differentially-rotating non-inertial frame centered on the
world-line $x^{\mu}(\tau )$ of an accelerated observer.

\bigskip

Moreover, it is shown in Ref.\cite{10a} that to each admissible
3+1 splitting are associated two congruences of time-like
observers (the natural ones for the given notion of simultaneity):

\noindent i) the Eulerian observers, whose unit 4-velocity field
is the field  of unit normals to the simultaneity surfaces
$\Sigma_{\tau}$;

\noindent ii) the observers whose unit 4-velocity field is
proportional to the evolution vector field of components $\partial
z^{\mu}(\tau ,\sigma^r)/\partial \tau$: in general this congruence
is non-surface forming having a non-vanishing vorticity (like the
congruence associated to a rotating disk).

\bigskip

The next problem is how to describe physical systems in
non-inertial frames and how to connect different conventions for
clock synchronization. The answer is given by {\it parametrized
Minkowski theories} \cite{12a}, \cite{1a}. Given any isolated
system (particles, strings, fields, fluids) admitting a Lagrangian
description, one makes the coupling of the system to an external
gravitational field and then replaces the 4-metric
${}^4g_{\mu\nu}(x)$ with the induced metric ${}^4g_{AB}[z(\tau
,\sigma^r)]$ associated to an arbitrary admissible 3+1 splitting.
The Lagrangian now depends not only on the matter configurational
variables but also on the embedding variables $z^{\mu}(\tau
,\sigma^r)$ (whose conjugate canonical momenta are denoted
$\rho_{\mu}(\tau ,\sigma^r)$). Since the action principle turns
out to be invariant under frame-preserving diffeomorphisms, at the
Hamiltonian level there are four first-class constraints ${\cal
H}_{\mu}(\tau ,\sigma^r) = \rho_{\mu}(\tau ,\sigma^r) -
l_{\mu}(\tau ,\sigma^r)\, T^{\tau\tau}(\tau ,\sigma^r) -
z^{\mu}_s(\tau ,\sigma^r)\, T^{\tau s}(\tau ,\sigma^r) \approx 0$
in strong involution with respect to Poisson brackets, $\{ {\cal
H}_{\mu}(\tau ,\sigma^r), {\cal H}_{\nu}(\tau ,\sigma_1^r)\} = 0$.
Here $l_{\mu}(\tau ,\sigma^r)$ are the covariant components of the
unit normal to $\Sigma_{\tau}$, while $z^{\mu}_s(\tau ,\sigma^r)$
are the components of three independent vectors tangent to
$\Sigma_{\tau}$. The quantities $T^{\tau\tau}$ and $T^{\tau s}$
are the components of the energy-momentum tensor of the matter
inside $\Sigma_{\tau}$ describing its energy- and momentum-
densities. As a consequence, Dirac's theory of constraints implies
that the configuration variables $z^{\mu}(\tau ,\sigma^r)$ are
arbitrary {\it gauge variables}. Therefore, all the admissible 3+1
splittings, namely all the admissible conventions for clock
synchronization, and all the admissible non-inertial frames
centered  on time-like observers are {\it gauge equivalent}.

By adding four gauge-fixing constraints $\chi^{\mu}(\tau
,\sigma^r) = z^{\mu}(\tau ,\sigma^r) - z^{\mu}_M(\tau ,\sigma^r)
\approx 0$ ($z^{\mu}_M(\tau ,\sigma^r)$ being an admissible
embedding), satisfying the orbit condition $det\,
|\{\chi^{\mu}(\tau ,\sigma^r), {\cal H}_{\nu}(\tau ,\sigma_1^r)|
\not= 0$, we identify the description of the system in the
associated non-inertial frame centered on a given time-like
observer. The resulting effective Hamiltonian for the
$\tau$-evolution turns out to contain the potentials of the {\it
relativistic inertial forces} present in the given non-inertial
frame. Since a non-inertial frame means the use of its radar
coordinates, we see that already in special relativity {\it
non-inertial Hamiltonians are coordinate-dependent quantities}
like the notion of energy density in general relativity.

\bigskip

As a consequence, the gauge variables $z^{\mu}(\tau ,\sigma^r)$
describe the {\it spatio-temporal appearances} of the phenomena in
non-inertial frames, which, in turn, are associated to extended
physical laboratories using a metrology for their measurements
compatible with the notion of simultaneity of the non-inertial
frame (think to the description of the Earth given by GPS).
Therefore, notwithstanding mathematics tends to use only
coordinate-independent notions, physical metrology forces us to
consider intrinsically coordinate-dependent quantities like the
non-inertial Hamiltonians. For instance, the motion of satellites
around the Earth is governed by a set of empirical coordinates
contained in the software of NASA computers: this is a {\it
metrological standard of space-time around the Earth} with a
poorly understood connection with the purely theoretical
coordinate systems. In a few years the European Space Agency will
start the project ACES about the synchronization of a
high-precision laser-cooled atomic clock on the space station with
similar clocks on the Earth  surface by means of microwave
signals. If the accuracy of 5 picosec. will be achieved, it will
be possible to make a coordinate-dependent test of effects at the
order $1/c^3$, like the second order Sagnac effect (sensible to
Earth rotational acceleration) and the general relativistic
Shapiro time-delay created by the geoid \cite{13a}. The  one-way
velocity of light between an Earth station and the space station
and the synchronization of the respective clocks are two faces of
the same problem.

\bigskip

Inertial frames centered on inertial observers are a special case
of gauge fixing in parametrized Minkowski theories. For each
configuration of an isolated system there is an special 3+1
splitting associated to it: the foliation with space-like
hyper-planes orthogonal to the conserved time-like 4-momentum of
the isolated system. This identifies an intrinsic inertial frame,
the {\it rest-frame}, centered on a suitable inertial observer
(the Fokker-Pryce center of inertia of the isolated system) and
allows to define the {\it Wigner-covariant rest-frame instant form
of dynamics} for every isolated system (see Ref.\cite{14a} for the
various forms of dynamics).

\bigskip

Let us remark that in parametrized Minkowski theories a
relativistic particle with world-line $x^{\mu}_i(\tau )$ is
described only by the 3-coordinates $\sigma^r = \eta^r_i(\tau )$
defined by $x^{\mu}_i(\tau ) = z^{\mu}(\tau , \eta^r_i(\tau ))$
and by the conjugate canonical momenta $\kappa_{ir}(\tau )$. The
usual 4-momentum $p_{i\mu}(\tau )$ is a derived quantity
satisfying the mass-shell constraint $\epsilon\, p^2_i = m^2_i$.
Therefore, we have a different description for positive- and
negative- energy particles. All the particles on an admissible
surface $\Sigma_{\tau}$ are simultaneous by construction: {\it
this eliminates the problem of relative times}, which for a long
time has been an obstruction to the theory of relativistic bound
states and to relativistic statistical mechanics (see
Ref.\cite{12a} and its bibliography for these problems and the
related no-interaction theorem).

\bigskip

Let us also remark that, differently from Fermi coordinates (a
purely theoretical construction), radar 4-coordinates can be {\it
operationally} defined. As shown in Ref.\cite{10a}, given four
functions satisfying certain restrictions induced by the M$\o$ller
conditions, the on-board computer of a spacecraft may establish a
grid of radar 4-coordinates in its future.

\bigskip

The discovery of the rest-frame instant form made possible to
develop a coherent formalism for all the aspects of relativistic
kinematics both for N particle systems and continuous bodies and
fields \cite{15a,16a} generalizing all known non-relativistic
results \cite{5a}:

\noindent i) the classification of the intrinsic notions of
collective variables (canonical non-covariant center of mass;
covariant non-canonical Fokker-Pryce center of inertia;
non-covariant non-canonical M$\o$ller center of energy);

\noindent ii) canonical bases of center-of-mass and relative
variables;

\noindent iii) canonical spin bases and dynamical body-frames for
the rotational kinematics of deformable systems;

\noindent iv) multipolar expansions for isolated and open systems;

\noindent v) the relativistic theory of orbits (while the
potentials appearing in the energy generator of the Poincare'
group determine the relative motion, the determination of the
actual orbits in the given inertial frame is influenced by the
potentials appearing in the Lorentz boosts: the vanishing of the
boosts is the natural gauge fixing to the rest-frame conditions
and selects the covariant Fokker-Pryce center of inertia);

\noindent vi) the M$\o$ller radius (a classical unit of length
identifying the region of non-covariance of the canonical center
of  mass of a spinning system around the covariant Fokker-Pryce
center of inertia; it is an effect induced by the Lorentz
signature of the 4-metric; it could be used as a physical
ultraviolet cutoff in quantization).

See Ref. \cite{17a} for a comprehensive review.
\bigskip

All these developments relied on an accurate study of the
structure of the constraint manifold $\bar \gamma$ (see the
Appendix) from the point of view of the orbits of the Poincare'
group. If $p^{\mu}$ is the total momentum of the system, the
constraint manifold has to be divided in four strata (some of them
may be absent for certain systems) according to whether $\sgn\,
p^2 > 0$, $p^2=0$, $\sgn\, p^2 < 0$ or $p^{\mu}=0$. Due to the
different little groups of the various Poincare' orbits, the gauge
orbits of different sectors will not be diffeomorphic. Therefore
the manifold $\bar \gamma$ is a {\it stratified} manifold and the
gauge foliations of relativistic systems are nearly never nice,
but rather one has to do with {\it singular} foliations.

For an acceptable relativistic system the stratum $\sgn\, p^2 < 0$
has to be absent to avoid tachyons. To study the strata $p^2=0$
and $p^{\mu}=0$ one has to add these relations as extra
constraints. For all the strata the next step is to do a canonical
transformation from the original variables to a new set consisting
of center-of-mass variables $x^{\mu}$, $p^{\mu}$ and of variables
relative to the center of mass, both for particle and field
systems. Let us consider the stratum $\sgn\, p^2 > 0$. By using
the standard Wigner boost $L^{\mu}_{\nu}(p, {\buildrel \circ \over
p})$ ($p^{\mu}=L^{\mu}_{\nu}(p,{\buildrel \circ \over
p}){\buildrel \circ \over p}^{\nu}$, ${\buildrel \circ \over
p}^{\mu}=\eta \sqrt {\sgn\, p^2} (1;\vec 0 )$, $\eta = sign\,
p^o$), one boosts the relative variables at rest. The new
variables are still canonical and the base is completed by
$p^{\mu}$ and by a new canonical non-covariant \footnote{This is
{\it a universal breaking of manifest Lorentz covariance}, which
appears for every isolated relativistic system, when one wants to
identify a canonical 4-center of mass. Since there is no
definition of relativistic center of mass enjoying all the
properties of the non-relativistic one, it turns out \cite{17a}
that the canonical 4-center of mass has only O(3) covariance like
the quantum Newton-Wigner position operator.} 4-center-of-mass
coordinate ${\tilde x}^{\mu}$, differing from $x^{\mu}$ for spin
terms. The new relative variables are either Poincare' scalars or
Wigner spin-1 vectors, transforming under the group O(3)(p) of the
Wigner rotations induced by the Lorentz transformations. A final
canonical transformation \cite{18a}, leaving fixed the relative
variables, sends the center-of-mass coordinates ${\tilde
x}^{\mu}$, $p^{\mu}$ in the new set $p\cdot {\tilde x}/\eta \sqrt
{\sgn\, p^2}=p\cdot x/\eta \sqrt {\sgn\, p^2}$ (the time in the
rest frame), $\eta \sqrt {\sgn\, p^2}$ (the total mass), $\vec k
=\vec p /\eta \sqrt {\sgn\, p^2}$ (the spatial components of the
unit 4-velocity $k^{\mu}= p^{\mu}/\eta \sqrt {\sgn\, p^2}$,
$k^2=1$), $\vec z=\eta \sqrt {\sgn\, p^2}( {\vec {\tilde
x}}-{\tilde x}^o\vec p/p^o)$. $\vec z$ is a non-covariant
3-center-of-mass canonical coordinate multiplied by the total
mass: it is the classical analog of the Newton-Wigner position
operator (like it, $\vec z$ is covariant only under the little
group O(3)(p) of the time-like Poincar\'e orbits). Analogous
considerations could be done for the other sectors. In Refs.
\cite{19a} there is the definition of other canonical bases, the
spin bases, adapted to the spin Casimir of the Poincar\'e group,
which made possible the quoted developments \cite{15a,16a,17a},
\cite{5a}.

\medskip

The nature of the relative variables depends on the system. The
first class constraints, once rewritten in terms of the new
variables, can be manipulated to find suitable global and Lorentz
scalar Abelianizations by means of Shanmugadhasan canonical
transformations (see the Appendix). Usually there is a combination
of the constraints which determines $\eta \sqrt {\sgn\, p^2}$,
i.e. the mass spectrum, so that the time in the rest frame $p\cdot
x/\eta \sqrt {\sgn\, p^2}$ is the conjugated Lorentz scalar gauge
variable. The other first class constraints eliminate some of the
relative variables \footnote{In particular they eliminate the {\it
relative energies} for systems of interacting relativistic
particles (and for the string), so that {\it the conjugate
relative times are gauge variables}, describing the freedom of the
observer of looking at the particle at the same time in every
allowed non-inertial frame \cite{12a}.}: their conjugated
coordinates are the other gauge variables. The DO (apart from the
center-of-mass ones $\vec k$ and $\vec z$ describing a decoupled
non-covariant observer) have to be extracted from the remaining
relative variables and the construction shows that they will be
either Poincare' scalars or Wigner covariant objects. In this way
in each stratum preferred global Shanmugadhasan canonical
transformations are identified, when no other kind of obstruction
to globality is present inside the various strata. \bigskip

See Ref. \cite{1a} for a list of the finite-dimensional
relativistic particle systems, which have been described with
Dirac theory of constraints and with parametrized Minkowski
theories, with some comments on their quantization.

\medskip

In particular in Ref.\cite{20a} there is the quantization of
relativistic scalar and spinning particles in a class of
non-inertial frames, whose simultaneity surfaces $\Sigma_{\tau}$
are space-like hyper-planes with arbitrary admissible linear
acceleration and carrying arbitrary admissible differentially
rotating 3-coordinates. It is based on a multi-temporal
quantization scheme for systems with first-class constraints, in
which only the particle degrees of freedom $\eta^r_i(\tau )$,
$\kappa_{ir}(\tau )$ are quantized. The gauge variables,
describing the appearances (inertial effects) of the motion in
non-inertial frames, are treated as c-numbers (like the time in
the Schroedinger equation with a time-dependent Hamiltonian) and
the physical scalar product does not depend on them. The
previously quoted relativistic kinematics has made possible to
separate the center of mass \footnote{At the relativistic level
this is done with a canonical transformation which is {\it point
only} in the momenta \cite{17a}.} and to verify that the spectra
of relativistic bound states in non-inertial frames are only
modified by inertial effects, being obtained from the inertial
ones by means of a time-dependent unitary transformation. The
non-relativistic limit \cite{9a} allows to recover the few
existing attempts of quantization in non-inertial frames as
particular cases.

\bigskip

Let us now look at what is known for other physically relevant
systems. In non-inertial frames there is the reformulation of the
Klein-Gordon equation \cite{21a}, of the Dirac equation
\cite{22a}, of the electro-magnetic field \cite{12a} and of
relativistic fluids \cite{23a} by means of parametrized Minkowski
theories.

Inspired by Dirac \footnote{Dirac \cite{24a} showed that the DO of
the electromagnetic field are the transverse vector potential
${\vec A}_{\perp}$ and the transverse electric field ${\vec
E}_{\perp}$ of the {\it radiation gauge}. When a fermion field is
interacting with the electromagnetic field, the fermionic DO is
{\it a fermion field dressed with a Coulomb cloud}.}, the
canonical reduction to Wigner-covariant generalized radiation
gauges, with the determination of the physical Hamiltonian as a
function of a canonical basis of DO, has been achieved for the
following isolated systems \footnote{For them one only asks that
the 10 conserved generators of the Poincar\'e algebra are finite
so to be able to use group theory; theories with external fields
can only be recovered as limits in some parameter of a subsystem
of the isolated system.}:

1) In the semi-classical approximation we get a consistent
description of N charged particles plus the electromagnetic field
if we use Grassmann -valued electric charges to regularize the
Coulomb self-energies. In Ref.\cite{25a} the electromagnetic
degrees of freedom are expressed in terms of the particle
variables by means the Lienard-Wiechert solution and this allows
to find the relativistic Darwin potential (or the Salpeter
potential for spinning particles) starting from classical
electrodynamics and not as a reduction from QFT.

2) Both the open and closed Nambu string, after an initial study
with light-cone coordinates, have been treated \cite{26a} along
these lines in the stratum $\sgn\, p^2 > 0$. Both Abelian Lorentz
scalar constraints and gauge variables have been found and
globally decoupled, and a redundant set of DO $[\vec z,\vec
k,{\vec {\tilde a}}_n]$ has been found. It remains an open problem
whether one can extract a global canonical basis of DO from the
Wigner spin 1 vectors ${\vec {\tilde a}}_n$, which satisfy
sigma-model-like constraints; if this basis exists, it would
define the Liouville integrability of the Nambu string and would
clarify whether there is any way to quantize it in four
dimensions.

3) Yang-Mills theory with Grassmann-valued fermion fields
\cite{27a} in the case of a trivial principal bundle over a
fixed-$x^o$ $R^3$ slice of Minkowski space-time with suitable
Hamiltonian-oriented boundary conditions; this excludes monopole
solutions and, since $R^3$ is not compactified, one has only
winding number and no instanton number. After a discussion of the
Hamiltonian formulation of Yang-Mills theory, of its group of
gauge transformations and of the Gribov ambiguity (see the
Appendix), the theory has been studied in suitable weighted
Sobolev spaces where the Gribov ambiguity is absent \cite{28a,29a}
and the global color charges are well defined. The global DO are
the transverse quantities ${\vec A}_{a\perp} (\vec x,x^o)$, ${\vec
E}_{a\perp}(\vec x,x^o)$ and fermion fields dressed with
Yang-Mills (gluonic) clouds. The nonlocal and non-polynomial (due
to the presence of classical Wilson lines along flat geodesics)
physical Hamiltonian has been obtained: it is nonlocal but without
any kind of singularities, it has the correct Abelian limit if the
structure constants are turned off, and it contains the explicit
realization of the abstract Mitter-Viallet metric.

4) SU(3) Yang-Mills theory with scalar particles with
Grassmann-valued color charges \cite{30a} for the regularization
of self-energies. It is possible to show that in this relativistic
scalar quark model the Dirac Hamiltonian expressed as a function
of DO has the property of asymptotic freedom.

5) The Abelian and non-Abelian SU(2) Higgs models with fermion
fields \cite{31a}, where the symplectic decoupling is a refinement
of the concept of unitary gauge. There is an ambiguity in the
solutions of the Gauss law constraints, which reflects the
existence of disjoint sectors of solutions of the Euler-Lagrange
equations of Higgs models. The physical Hamiltonian and Lagrangian
of  the Higgs phase have been found; the self-energy turns out to
be local and contains a local four-fermion interaction.

6) The standard SU(3)xSU(2)xU(1) model of elementary particles
\cite{32a} with \hfill\break Grassmann- valued fermion fields. The
final reduced Hamiltonian contains nonlocal self-energies for the
electromagnetic and color interactions, but ``local ones" for the
weak interactions implying the non-perturbative emergence of
4-fermions interactions.

\bigskip

In inertial frames the quantization of DO can be faced with the
standard methods.  The main open problem is the quantization of
the scalar Klein-Gordon field in non-inertial frames, due to the
Torre and Varadarajan \cite{33a} no-go theorem, according to which
in general the evolution from an initial space-like hyper-surface
to a final one is {\it  not unitary} in the Tomonaga-Schwinger
formulation of quantum field theory. From the 3+1 point of view
there is evolution only among the leaves of an admissible
foliation and the possible way out from the theorem lies in the
determination of all the admissible 3+1 splittings of Minkowski
space-time satisfying the following requirements: i) existence of
an instantaneous Fock space on each simultaneity surface
$\Sigma_{\tau}$ (i.e. the $\Sigma_{\tau}$'s must admit a
generalized Fourier transform); ii) unitary equivalence of the
Fock spaces on $\Sigma_{\tau_1}$ and $\Sigma_{\tau_2}$ belonging
to the same foliation (the associated Bogoljubov transformation
must be Hilbert-Schmidt), so that the non-inertial Hamiltonian is
a Hermitean operator; iii) unitary gauge equivalence of the 3+1
splittings with the Hilbert-Schmidt property. The overcoming of
the no-go theorem would help also in quantum field theory in
curved space-times and in condensed matter (here the non-unitarity
implies non-Hermitean Hamiltonians and negative energies).

\section{The Dynamical Chrono-Geometrical Structure of General
Relativity: the Rest-Frame Instant Form of Metric and Tetrad
Gravity and the Role of Non-Inertial Frames.}

In the years 1913-16 Einstein developed general relativity relying
on the {\it equivalence principle} (equality of inertial and
gravitational masses of bodies in free fall). It suggested him the
impossibility to distinguish a uniform gravitational field from
the effects of a constant acceleration by  means of local
experiments in sufficiently small regions where the effects of
tidal forces are negligible. This led to the {\it geometrization}
of the gravitational interaction and to the replacement of
Minkowski space-time with a pseudo-Riemannian 4-manifold $M^4$
with non vanishing curvature Riemann tensor. The principle of {\it
general covariance} (see Ref.\cite{34a} for a review), at the
basis of the tensorial nature of Einstein's equations, has the two
following consequences:

i) the invariance of the Hilbert action under {\it passive}
diffeomorphisms (the coordinate transformations in $M^4$), so that
the second Noether theorem implies the existence of first-class
constraints at the Hamiltonian level;

ii) the mapping of solutions of Einstein's equations among
themselves under the action of {\it active} diffeomorphisms of
$M^4$ extended to the tensors over $M^4$ (dynamical symmetries of
Einstein's equations).

\bigskip

The basic field of metric gravity is the 4-metric tensor with
components ${}^4g_{\mu\nu}(x)$ in an arbitrary coordinate system
of $M^4$. The peculiarity of gravity is that the 4-metric field,
differently from the fields of electromagnetic, weak and strong
interactions and from the matter fields, has a {\it double role}:

i) it is the mediator of the gravitational interaction (in analogy
to all the other gauge fields);

ii) it determines the chrono-geometric structure of the space-time
$M^4$ in a dynamical way through the line element $ds^2 =
{}^4g_{\mu\nu}(x)\, dx^{\mu}\, dx^{\nu}$.

As a consequence, the gravitational field {\it teaches
relativistic causality} to all the other fields: for instance it
tells to classical rays of light and to quantum photons and gluons
which are the allowed trajectories for massless particles in each
point of $M^4$.

\bigskip

Let us make a comment about the two main existing approaches to
the quantization of gravity.

\medskip

1) {\it Effective quantum field theory and string theory}. This
approach contains the standard model of elementary particles and
its extensions. However, since the quantization, namely the
definition of the Fock space, requires a background space-time
where it is possible to define creation and annihilation
operators, one must use the splitting ${}^4g_{\mu\nu} =
{}^4\eta^{(B)}_{\mu\nu} + {}^4h_{\mu\nu}$ and quantize only the
perturbation ${}^4h_{\mu\nu}$ of the background 4-metric
$\eta^{(B)}_{\mu\nu}$ (usually $B$ is either Minkowski or DeSitter
space-time). In this way property ii) is lost (one uses the fixed
non-dynamical chrono-geometrical structure of the background
space-time), gravity is replaced by a   field of spin two over the
background (and passive diffeomorphisms are replaced by gauge
transformations acting in an inner space) and the only difference
among gravitons, photons and gluons lies in their quantum numbers.

\medskip

2) {\it Loop quantum gravity}. This approach never introduces a
background space-time, but being inequivalent to a Fock space, has
problems to incorporate particle physics. It uses a fixed 3+1
splitting of the space-time $M^4$ and it is a quantization of the
associated instantaneous 3-spaces $\Sigma_{\tau}$ (quantum
geometry). However, there is no known way to implement a
consistent unitary evolution (the problem of the super-hamiltonian
constraint) and, since it is usually formulated in spatially
compact space-times without boundary, there is no notion of a
Poincare' group (and therefore no extra dimensions) and a problem
of time (frozen picture without evolution).

\bigskip

For  outside points of view on loop quantum gravity  and string
theory see Refs. \cite{35a,36a}, respectively.

\bigskip

Let us remark that in all known formulations particle and nuclear
physics are a chapter of the theory of representations of the
Poincare' group in inertial frames in the spatially non-compact
Minkowski space-time. As a consequence, if one looks at general
relativity from the point of view of particle physics, the main
problem to get a unified theory is how to reconcile the Poincare'
group (the kinematical group of the transformations  connecting
inertial frames) with the diffeomorphism group implying the
non-existence of global inertial frames in general relativity
(special relativity holds only in a small neighborhood of a body
in free fall).

\bigskip

Let us consider the ADM formulation of metric gravity \cite{37a}
and its extension to tetrad gravity \footnote{It is needed to
describe the coupling of gravity to fermions; it is a theory of
time-like observers endowed with a tetrad field, whose time-like
axis is the unit 4-velocity of the observer and whose spatial axes
are associated to a choice of three gyroscopes.} obtained by
replacing the ten configurational 4-metric variables
${}^4g_{\mu\nu}(x)$ with the sixteen cotetrad fields
${}^4E^{(\alpha )}_{\mu}(x)$ by means of the decomposition
${}^4g_{\mu\nu}(x) = {}^4E^{(\alpha )}_{\mu}(x)\,
{}^4\eta_{(\alpha )(\beta )}\, {}^4E^{(\beta )}_{\nu}(x)$
[$(\alpha )$ are flat indices].

Then, after having restricted the model to globally hyperbolic,
topologically trivial, spatially non-compact space-times
(admitting a global notion  of time), let us introduce a global
3+1 splitting of the space-time $M^4$ and let choose the
world-line of a time-like observer. As in special relativity, let
us make a coordinate transformation to observer-dependent radar
4-coordinates, $x^{\mu} \mapsto \sigma^A = (\tau ,\sigma^r)$,
adapted to the 3+1 splitting and using the observer world-line as
origin of the 3-coordinates. Again the inverse transformation,
$\sigma^A \mapsto x^{\mu} = z^{\mu}(\tau ,\sigma^r)$, defines the
embedding of the leaves $\Sigma_{\tau}$ into $M^4$. These leaves
$\Sigma_{\tau}$ (assumed to be Riemannian 3-manifolds
diffeomorphic to $R^3$, so that they admit global 3-coordinates
$\sigma^r$ and a unique 3-geodesic joining any pair of points in
$\Sigma_{\tau}$) are both Cauchy surfaces and simultaneity
surfaces corresponding to a convention for clock synchronization.
For the induced 4-metric we get

\begin{eqnarray*}
 {}^4g_{AB}(\sigma ) &=& {{\partial z^{\mu}(\sigma )}\over
{\partial \sigma^A}}\, {}^4g_{\mu\nu}(x)\, {{\partial
z^{\nu}(\sigma )}\over {\partial \sigma^B}} =
 {}^4E^{(\alpha )}_A\, {}^4\eta_{(\alpha )(\beta )}\,
{}^4E^{(\beta )}_B = \nonumber \\
 &=&\epsilon \left( \begin{array}{cc} (N^2- {}^3g_{rs}\, N^r\, N^s) &
- {}^3g_{su}\, N^u\\ - {}^3g_{ru}\, N^u & -{}^3g _{rs} \end{array}
\right)(\sigma ).\nonumber \\
 \end{eqnarray*}

Here ${}^4E^{(\alpha )}_A(\tau ,\sigma^r)$ are adapted cotetrad
fields, $N(\tau ,\sigma^r)$ and $N^r(\tau ,\sigma^r)$ the lapse
and shift functions and ${}^3g_{rs}(\tau ,\sigma^r)$ the 3-metric
on $\Sigma_{\tau}$ with signature $(+ + +)$. We see that in
general relativity the quantities $z^{\mu}_A = \partial
z^{\mu}/\partial \sigma^A$  are no more cotetrad fields on $M^4$
differently from what happens in special relativity: now they  are
only transition functions between coordinate charts, so that  the
dynamical fields are now the real cotetrad fields ${}^4E^{(\alpha
)}_A(\tau ,\sigma^r)$ and not the embeddings $z^{\mu}(\tau
,\sigma^r)$.

\bigskip

Let us try to identify a class of space-times and an associated
suitable family of admissible 3+1 splittings able to incorporate
particle physics and giving a model for the solar system or our
galaxy (and hopefully allowing an extension to the cosmological
context) with the following further requirements \cite{38a}:
\medskip

1) $M^4$ must be asymptotically flat at spatial infinity and the
4-metric must  tend asymptotically at spatial infinity to the
Minkowski 4-metric in every coordinate system (this implies that
the 4-diffeomorphisms must tend to the identity at spatial
infinity). Therefore, in these space-times there is an {\it
asymptotic background 4-metric} and this will allow to avoid the
decomposition ${}^4g_{\mu\nu} = {}^4\eta_{\mu\nu} +
{}^4h_{\mu\nu}$ in the bulk.
\medskip

2) The boundary conditions on the fields on each leaf
$\Sigma_{\tau}$ of the admissible 3+1 splittings must be such to
reduce the Spi group of asymptotic symmetries (see Ref.\cite{39a})
to the ADM Poincare' group. This means that {\it
super-translations} (direction-dependent quasi Killing vectors,
obstruction to the definition of angular momentum in general
relativity) must be absent, namely that all the fields must tend
to their asymptotic limits in a direction- independent way (see
Refs. \cite{40a}). This is possible only if the admissible 3+1
splittings have all the leaves $\Sigma_{\tau}$ tending to
Minkowski space-like hyper-planes orthogonal to the ADM 4-momentum
at spatial infinity \cite{38a}. In turn this implies that every
$\Sigma_{\tau}$ is {\it the rest frame of the instantaneous
3-universe} and that there are asymptotic inertial observers to be
identified with the {\it fixed stars} \footnote{In a future
extension to the cosmological context they could be identified
with the privileged observers at rest with respect to the
background cosmic radiation.}. This requirement implies that the
shift functions vanish at spatial infinity [$N^r(\tau ,\sigma^r)\,
\rightarrow O(1/|\sigma |^\epsilon )$, $\epsilon > 0$, $\sigma^r =
|\sigma |\, {\hat u}^r$], where the lapse function tends to $1$
[$N(\tau ,\sigma^r)\, \rightarrow\, 1 + O(1/|\sigma |^\epsilon )$]
and the 3-metric tends to the Euclidean one [${}^3g_{rs}(\tau
,\sigma^u)\, \rightarrow\, \delta_{rs} + O(1/|\sigma |)$].
\medskip

3) The admissible 3+1 splittings should have the leaves
$\Sigma_{\tau}$ admitting a generalized Fourier transform (namely
they should be Lichnerowicz \cite{41a} 3-manifolds with
involution, so to have the possibility to define instantaneous
Fock spaces in a future attempt of quantization).

\medskip

4) All the fields on $\Sigma_{\tau}$ should belong to suitable
weighted Sobolev spaces, so that $M^4$ has no Killing vectors and
Yang-Mills fields on $\Sigma_{\tau}$ do not present Gribov
ambiguities (due to the presence of gauge symmetries and gauge
copies) \cite{27a}.
\bigskip

In absence of matter the Christodoulou and Klainermann \cite{42a}
space-times are good candidates: they are near Minkowski
space-time in a norm sense, avoid singularity theorems by relaxing
the requirement of conformal completability (so that it is
possible to follow solutions of Einstein's equations on long
times) and admit gravitational radiation at null infinity.

\bigskip

Since the simultaneity leaves $\Sigma_{\tau}$ are the rest frame
of the instantaneous 3-universe,  at the Hamiltonian level it is
possible to define {\it the rest-frame instant form of metric and
tetrad gravity} \cite{38a,43a}. If matters is present, the limit
of this description for vanishing Newton constant will produce the
rest-frame instant form description of the same matter in the
framework of parametrized Minkowski theories and the ADM Poincare'
generators will tend to the kinematical Poincare' generators of
special relativity. Therefore we have obtained a model admitting
{\it a deparametrization of general relativity to special
relativity}. It is not known whether the rest-frame condition can
be relaxed in general relativity without having super-translations
reappearing, since the answer to this question is connected with
the non-trivial problem of boosts in general relativity.

\bigskip

Let us now come back to ADM tetrad gravity. The time-like vector
${}^4E^A_{(o)}(\tau ,\sigma^r)$ of the tetrad field
${}^4E^A_{(\alpha )}(\tau ,\sigma^r)$, dual to the cotetrad field
${}^4E^{(\alpha )}_A(\tau ,\sigma^r)$, may be rotated to become
the unit normal to $\Sigma_{\tau}$ in each point by means of a
standard Wigner boost for time-like Poincare' orbits depending on
three parameters $\varphi_{(a)}(\tau ,\sigma^r)$, $a = 1,2,3$:
${}^4E^A_{(o)}(\tau ,\sigma^r) = L^A{}_B(\varphi_{(a)}(\tau
,\sigma^r))\, {}^4{\check E}^B_{(o)}(\tau ,\sigma^r)$. This allows
to define the following cotetrads adapted to the 3+1 splitting
(the so-called {\it Schwinger time gauge}) ${}^4{\check
E}^{(o)}_A(\tau ,\sigma^r) = \Big(N(\tau ,\sigma^r); 0\Big)$,
${}^4{\check E}^{(a)}_A(\tau ,\sigma^r) = \Big(N_{(a)}(\tau
,\sigma^r); {}^3e_{(a)r}(\tau ,\sigma^r)\Big)$, where
${}^3e_{(a)r}(\tau ,\sigma^r)$ are cotriads fields on
$\Sigma_{\tau}$ (tending to $\delta_{(a)r} + O(1/|\sigma |)$ at
spatial infinity) and $N_{(a)} = N^r\, {}^3e_{(a)r}$. As a
consequence, the sixteen cotetrad fields may be replaced by the
fields $\varphi_{(a)}(\tau ,\sigma^r)$, $N(\tau ,\sigma^r)$,
$N_{(a)}(\tau ,\sigma^r)$, ${}^3e_{(a)r}(\tau ,\sigma^r)$, whose
conjugate canonical momenta will be denoted as $\pi_N(\tau
,\sigma^r)$, $\pi_{\vec N\, (a)}(\tau ,\sigma^r)$, $\pi_{\vec
\varphi\, (a)}(\tau ,\sigma^r)$, ${}^3\pi^r_{(a)}(\tau
,\sigma^r)$.

\bigskip

The local invariance of the ADM action imply the existence of 14
first-class constraints (10 primary and 4 secondary):

i) $\pi_N(\tau ,\sigma^r) \approx 0$ implying the secondary
super-hamiltonian constraint ${\cal H}(\tau ,\sigma^r) \approx 0$;

ii) $\pi_{\vec N\, (a)}(\tau ,\sigma^r) \approx 0$ implying the
secondary super-momentum constraints ${\cal H}_{(a)}(\tau
,\sigma^r) \approx 0$;

iii) $\pi_{\vec \varphi\, (a)}(\tau ,\sigma^r) \approx 0$;

iv) three constraints $M_{(a)}(\tau ,\sigma^r) \approx 0$
generating rotations of the cotriads.

As a consequence there are 14 gauge variables describing the {\it
generalized inertial effects} in the non-inertial frame defined by
the chosen admissible 3+1  splitting of $M^4$ centered on an
arbitrary time-like observer. The remaining independent "two +
two" degrees of freedom are the gauge invariant DO of the
gravitational field describing {\it generalized tidal effects}.
The same degrees  of freedom emerge in ADM metric gravity, where
the configuration variables $N$, $N^r$, ${}^4g_{rs}$ with
conjugate momenta $\pi_N$, $\pi_{\vec N\, r}$, ${}^3\Pi^{rs}$, are
restricted by 8 first-class constraints ($\pi_N(\tau ,\sigma^r)
\approx 0\, \rightarrow {\cal H}(\tau ,\sigma^r) \approx 0$,
$\pi_{\vec N\, r}(\tau ,\sigma^r) \approx 0 \, \rightarrow\, {\cal
H}^r(\tau ,\sigma^r) \approx 0$).
\bigskip

Again it is possible to make a separation of the gauge variables
from the DO by means of a Shanmugadhasan canonical transformation.
Since no-one knows how to solve the super-hamiltonian constraint
(except that in the post-Newtonian approximation), the best we can
do is to look for a quasi-Shanmugadhasan canonical transformation
adapted to the other 13 first-class constraints (the only
constraints to be Abelianized are $M_{(a)}(\tau ,\sigma^r) \approx
0$ and ${\cal H}_{(a)}(\tau ,\sigma^r) \approx 0$) \cite{43a}:

\begin{equation}
\begin{minipage}[t]{3cm}
\begin{tabular}{|l|l|l|l|} \hline
$\varphi^{(a)}$ & $N$ & $N_r$ & ${}^3e_{(a)r}$ \\ \hline $\approx
0$ & $\approx 0$ & $  \approx 0 $ & ${}^3{\tilde \pi}^r_{(a)}$
\\ \hline
\end{tabular}
\end{minipage} \hspace{1cm} {\longrightarrow \hspace{.2cm}} \
\begin{minipage}[t]{4 cm}
\begin{tabular}{|lllll|l|l|} \hline
$\varphi^{(a)}$ & $N$ & $N_{(a)}$ & $\alpha_{(a)}$ & $\xi^{r}$ &
$\phi$ & $r_{\bar a}$\\ \hline $\approx0$ &
 $\approx 0$ & $\approx 0$ & $\approx 0$
& $\approx 0$ &
 $\pi_{\phi}$ & $\pi_{\bar a}$ \\ \hline
\end{tabular}
\end{minipage}.
 \nonumber \\
 \end{equation}

Here, $\alpha_{(a)}(\tau ,\sigma^r)$ are three Euler angles and
$\xi^r(\tau ,\sigma^r)$ are three parameters giving a
coordinatization of the action of 3-diffeomorphisms on the
cotriads ${}^3e_{(a)r}(\tau ,\sigma^r)$. The configuration
variable $\phi (\tau ,\sigma^r) = \Big(det\, {}^3g(\tau
,\sigma^r)\Big)^{1/12}$ is the conformal factor of the 3-metric:
it can be shown that it is the unknown in the super-hamiltonian
constraint (also named the Lichnerowicz equation). The gauge
variables are $N$, $N_{(a)}$, $\varphi_{(a)}$, $\alpha_{(a)}$,
$\xi^r$ and $\pi_{\phi}$, while $r_{\bar a}$, $\pi_{\bar a}$,
$\bar a = 1,2$, are the DO of the gravitational field (in general
they are not tensorial quantities).

\bigskip

Even if we do not know the expression of the final variables in
terms of the original ones, we note that this is a {\it point}
canonical transformation with known inverse

\beq
 {}^3e_{(a)r}(\tau ,\sigma^u ) =
 {}^3R_{(a)(b)}(\alpha_{(e)}(\tau ,\sigma^u ))\, {{\partial
 \xi^s(\tau ,\sigma^u )}\over {\partial \sigma^r}}\,
\phi^2(\tau , \vec \xi (\tau ,\sigma^u ))\,
 {}^3{\hat e}_{(b)s}( r_{\bar a}
(\tau , \xi^u (\tau ,\sigma^v ))\, ),\nonumber \\
\eeq

\noindent as implied by the study of the gauge transformations
generated by the first-class constraints (${}^3{\hat e}_{(a)r}$
are reduced cotriads, which depend only on the two configurational
DO $r_{\bar a}$).

\medskip

The point nature of the canonical transformation implies that the
old cotriad momenta are linear functionals of the new momenta. The
kernel connecting the old and new momenta satisfy elliptic partial
differential equations implied by i) the canonicity conditions;
ii) the super-momentum constraints ${\cal H}_{(a)}(\tau ,\sigma^r)
\approx 0$; iii) the rotation constraints $M_{(a)}(\tau ,\sigma^r)
\approx 0$.

\bigskip
As already said, the first-class constraints are the generators of
the Hamiltonian gauge transformations, under which the ADM action
is quasi-invariant (second Noether theorem):\medskip

i) The gauge transformations generated by the four primary
constraints $\pi_N(\tau ,\sigma^r) \approx 0$, $\pi_{\vec N\,
(a)}(\tau ,\sigma^r) \approx 0$, modify the lapse and shift
functions, namely how densely the simultaneity surfaces are packed
in $M^4$ and which points have the same 3-coordinates on each
$\Sigma_{\tau}$.

ii) Those generated by the three super-momentum constraints ${\cal
H}_{(a)}(\tau ,\sigma^r) \approx 0$ change the 3-coordinates on
$\Sigma_{\tau}$.

iii) Those generated by the super-hamiltonian constraint ${\cal
H}(\tau ,\sigma^r) \approx 0$ transform an admissible 3+1
splitting into another admissible one by realizing a normal
deformation of the simultaneity surfaces $\Sigma_{\tau}$
\cite{44a}. As a consequence, {\it all the conventions about clock
synchronization are gauge equivalent as in special relativity}.

iv) Those generated by $\pi_{\vec \varphi\, (a)}(\tau ,\sigma^r)
\approx 0$, $M_{(a)}(\tau ,\sigma^r) \approx 0$, change the
cotetrad fields with local Lorentz transformations.

\medskip

In the rest-frame instant form of tetrad gravity there are the
three extra first-class constraints $P^r_{ADM} \approx 0$
(vanishing of the ADM 3-momentum as {\it rest-frame conditions}).
They generate gauge transformations which change the time-like
observer whose world-line is used as origin of the 3-coordinates.

\bigskip

Finally let us see which is the Dirac Hamiltonian $H_D$ generating
the $\tau$-evolution in ADM canonical gravity. In {\it spatially
compact space-times without boundary} $H_D$ is a linear
combination of the primary constraints  plus the secondary
super-hamiltonian and super-momentum constraints multiplied by the
lapse and shift functions respectively (consequence of the
Legendre transform). As a consequence, $H_D \approx 0$ and in the
reduced phase space  we get a vanishing Hamiltonian. This implies
the so-called {\it frozen picture} and the problem of how to
reintroduce a temporal evolution \footnote{See Refs.\cite{45a} for
the problem of time in general relativity.}. Usually one considers
the normal (time-like) deformation of $\Sigma_{\tau}$ induced by
the super-hamiltonian constraint as an evolution in a local time
variable to be identified (the multi-fingered time point of view
with a local either extrinsic or intrinsic time): this is the
so-called {\it Wheeler-DeWitt interpretation} \footnote{Kuchar
\cite{46a} says that the super-hamiltonian constraint must not be
interpreted as a generator of gauge transformations, but as an
effective Hamiltonian.}.

\medskip

On the contrary, {\it in spatially non-compact space-times} the
definition of functional derivatives and the existence of a
well-posed Hamiltonian action principle (with the possibility of a
good control of the surface terms coming from integration by
parts) require the addition of the {\it DeWitt \cite{47a} surface
term} (living on the surface at spatial infinity) to the
Hamiltonian. It can be shown \cite{38a} that in the rest-frame
instant form this term, together with a surface term coming from
the Legendre transformation of the ADM action, leads to the Dirac
Hamiltonian

\beq
 H_D = {\check E}_{ADM} + (constraints) =
  E_{ADM} + (constraints) \approx E_{ADM}.\nonumber \\
 \eeq

\noindent Here ${\check E}_{ADM}$ is the {\it strong ADM energy},
a surface term analogous to the one defining the electric charge
as the flux of the electric field through the surface at spatial
infinity in electromagnetism. Since we have ${\check E}_{ADM} =
E_{ADM} + (constraints)$, we see that the non-vanishing part of
the Dirac Hamiltonian is the {\it weak ADM energy} $E_{ADM} = \int
d^3\sigma\, {\cal E}_{ADM}(\tau ,\sigma^r)$, namely the integral
over $\Sigma_{\tau}$ of the ADM energy density (in
electromagnetism this corresponds to the definition of the
electric charge as the volume integral of matter charge density).
Therefore there is no frozen picture but a consistent
$\tau$-evolution.
 \medskip

However, the ADM energy density ${\cal E}_{ADM}(\tau ,\sigma^r)$
is a {\it coordinate-dependent quantity}, because it depends on
the gauge variables (namely on the relativistic inertial effects
present in the non-inertial frame): this is the {\it problem of
energy} in general relativity. Let us remark that in most
coordinate systems ${\cal E}_{ADM}(\tau ,\sigma^r)$ does not agree
with the pseudo-energy density defined in terms of the
Landau-Lifschiz pseudo-tensor.

 \bigskip

As a consequence, to get a deterministic evolution for the DO
\footnote{See Refs.\cite{48a} for the modern formulation of the
Cauchy problem for Einstein equations, which mimics the steps of
the Hamiltonian formalism.} we must fix the gauge completely, that
is we have to add 14 gauge-fixing constraints satisfying an orbit
condition and to pass to Dirac brackets. As already said, the
correct way to do it is the following one:

i) Add a gauge-fixing constraint to the secondary
super-hamiltonian constraint \footnote{The special choice
$\pi_{\phi}(\tau ,\sigma^r) \approx 0$ implies that the DO
$r_{\bar a}$, $\pi_{\bar a}$, remain canonical even if we do not
know how to solve this constraint.}. This gauge-fixing fixes the
form of $\Sigma_{\tau}$, i.e. the convention for the
synchronization of clocks. The $\tau$-constancy of this
gauge-fixing constraint generates a gauge-fixing constraint to the
primary constraint $\pi_N(\tau ,\sigma^r) \approx 0$ for the
determination of the lapse function. The $\tau$-constancy of this
new gauge fixing determines the Dirac multiplier in front of the
primary constraint.

ii) Add three gauge-fixings to the secondary super-momentum
constraints ${\cal H}_{(a)}(\tau ,\sigma^r) \approx 0$. This fixes
the 3-coordinates on each $\Sigma_{\tau}$. The $\tau$-constancy of
these gauge fixings generates the three gauge fixings to the
primary constraints $\pi_{\vec N\, (a)}(\tau ,\sigma^r) \approx 0$
and leads to the determination of the shift functions (i.e. of the
appearances of gravito-magnetism). The $\tau$-constancy of these
new gauge fixings determines the Dirac multipliers in front of the
three primary constraints.

iii) Add six gauge-fixing constraints to the primary constraints
$\pi_{\vec \varphi\, (a)}(\tau ,\sigma^r) \approx 0$,
$M_{(a)}(\tau ,\sigma^r) \approx 0$. This is a fixation of the
cotetrad field which includes a convention on the choice and the
transport of the three gyroscopes of every time-like observer of
the two congruences associated to the chosen 3+1 splitting of
$M^4$. Their $\tau$-constancy determines the six Dirac multipliers
in front of these primary constraints.

iv) In the rest-frame instant form we must also add three gauge
fixings to the rest-frame conditions $P^r_{ADM} \approx 0$. The
natural ones are obtained with the requirement that the three ADM
boosts vanish. In this way we select a special time-like observer
as origin of the 3-coordinates (like the Fokker-Pryce center of
inertia in special relativity \cite{17a}).

\bigskip

In this way all the gauge variables are fixed to be either
numerical functions or well determined functions of the DO. As a
consequence, in a completely fixed gauge (i.e. in a non-inertial
frame centered on a time-like observer and with its pattern of
inertial forces, corresponding to an extended physical laboratory
with fixed metrological conventions) the ADM energy density ${\cal
E}_{ADM}(\tau ,\sigma^r)$ becomes {\it a well defined function
only of the DO} and the Hamilton equations for them with $E_{ADM}$
as Hamiltonian are a hyperbolic system of partial differential
equations for their determination. For each choice of Cauchy data
for the DO on a $\Sigma_{\tau}$, we obtain a solution of
Einstein's equations in the radar 4-coordinate system associated
to the chosen 3+1 splitting of $M^4$.
\medskip

A universe $M^4$ (a {\it 4-geometry}) is the equivalence class of
all the completely fixed gauges with gauge equivalent Cauchy data
for the DO on the associated Cauchy and simultaneity surfaces
$\Sigma_{\tau}$. In each gauge we find the solution for the DO in
that gauge ({\it the tidal effects}) and then the explicit form of
the gauge variables ({\it the inertial effects}). Moreover, {\it
also the extrinsic curvature of the simultaneity surfaces
$\Sigma_{\tau}$ is determined}. Since the simultaneity surfaces
are asymptotically flat, it is possible to determine their
embeddings $z^{\mu}(\tau ,\sigma^r)$ in $M^4$. As a consequence,
differently from special relativity, the conventions for clock
synchronization and the whole chrono-geometrical structure of
$M^4$ (gravito-magnetism, 3-geodesic  spatial distance on
$\Sigma_{\tau}$, trajectories of light rays in each point of
$M^4$, one-way velocity of light) are {\it dynamically determined
}.

\bigskip

Let us remark that, if we look at  Minkowski space-time as a
special solution of Einstein's equations with $r_{\bar a}(\tau
,\sigma^r) = \pi_{\bar a}(\tau ,\sigma^r) = 0$ (zero Riemann
tensor, no tidal effects, only inertial effects), we find
\cite{38a} that the dynamically admissible 3+1 splittings
(non-inertial frames) must have the simultaneity surfaces
$\Sigma_{\tau}$ {\it 3-conformally flat}, because the conditions
$r_{\bar a}(\tau ,\sigma^r) = \pi_{\bar a}(\tau ,\sigma^r) = 0$
imply the vanishing of the Cotton-York tensor of $\Sigma_{\tau}$.
Instead, in special relativity, considered as an autonomous
theory, all the non-inertial frames compatible with the M$\o$ller
conditions are admissible, namely there is much more freedom in
the conventions for clock synchronization.

\bigskip

A first application of this formalism \cite{49a} has been the
determination of post-Minkowskian background-independent
gravitational waves in a completely fixed non-harmonic
3-orthogonal gauge with diagonal 3-metric. It can be shown that
the requirements $r_{\bar a}(\tau ,\sigma^r) << 1$, $\pi_{\bar
a}(\tau ,\sigma^r) << 1$ lead to a weak field approximation based
on a Hamiltonian linearization scheme:

\noindent i) linearize the Lichnerowicz equation, determine the
conformal factor of the 3-metric and then the  lapse and shift
functions;

\noindent ii) find $E_{ADM}$ in this gauge and disregard all the
terms more than quadratic in the DO;

\noindent iii) solve the Hamilton equations for the DO.

In this way we get a solution of linearized Einstein's equations,
in which the configurational DO $r_{\bar a}(\tau ,\sigma^r)$ play
the role of the {\it two polarizations} of the gravitational wave
and we can evaluate the embedding $z^{\mu}(\tau ,\sigma^r)$ of the
simultaneity surfaces of this gauge explicitly.

\bigskip

Let us conclude with some remarks about the interpretation of the
space-time 4-manifold in general relativity.
\medskip

In 1914 Einstein, during his researches for developing general
relativity, faced the problem arising from the fact that the
requirement of general covariance would involve a threat to the
physical objectivity of the points of space-time $M^4$, which in
classical field theories are usually assumed to have a {\it well
defined individuality}. This led him to formulate the Hole
Argument. Assume that $M^4$ contains a {\it hole} ${\cal H}$, that
is an open region where all the non-gravitational fields vanish.
It  is implicitly assumed that the Cauchy surface for Einstein's
equations lies outside ${\cal H}$. Let us consider an active
diffeomorphism $A$ which re-maps the points inside ${\cal H}$, but
is the identity outside ${\cal H}$. For any point $x \in {\cal H}$
we have $x \mapsto D_A\, x \in {\cal H}$. The induced active
diffeomorphism on the 4-metric tensor ${}^4g$, solution of
Einstein's equations, will map it into another solution $D^*_A\,
{}^4g$ ($D^*_A$ is a dynamical symmetry of Einstein's equations)
defined by $D^*_A\, {}^4g(D_A\, x) = {}^4g(x) \not= D^*_A\,
{}^4g(x)$. As  a consequence, we get two solutions of Einstein's
equations with the same Cauchy data outside ${\cal H}$ and it is
not clear how to save the identification of the mathematical
points of $M^4$.

\bigskip

Einstein avoided the problem with the pragmatic {\it
point-coincidence argument}: the only real world-occurrences are
the (coordinate-independent) space-time coincidences (like the
intersection of two world-lines). However, the problem was
reopened by Stachel \cite{50a} and then by Earman and Norton
\cite{51a} and this opened a rich philosophical debate that is
still alive today.

If we insist on the reality of space-time mathematical points
independently from the presence of any physical field (the {\it
substantivalist} point of view in philosophy of science), we are
in trouble with predictability.

If we say that ${}^4g$ and $D^*_A\, {}^4g$ describe the same
universe (the so-called {\it Leibniz equivalence}), we loose any
physical objectivity of the space-time points (the {\it
relationist} point of view).

Stachel \cite{50a} suggested that a physical individuation of the
point-events of $M^4$ could be done only by using {\it four
individuating fields depending on the 4-metric on $M^4$}, namely
that a tensor field on $M^4$ is needed to identify the points of
$M^4$.
\medskip

On the other hand, {\it coordinatization} is the only way to
individuate the points {\it mathematically} since, as stressed by
Hermann Weyl \cite{52a}: ''There is no distinguishing objective
property by which one could tell apart one point from all others
in a homogeneous space: at this level, fixation of a point is
possible only by a {\it demonstrative act} as indicated by terms
like {\it this} and {\it there}.''

\bigskip

To clarify the situation let us remember that Bergmann and Komar
\cite{53a} gave {\it a passive re-interpretation of active
diffeomorphisms as metric-dependent coordinate transformations}
$x^{\mu} \mapsto y^{\mu}(x, {}^4g(x))$ restricted to the solutions
of Einstein's equations (i.e. {\it on-shell}). It can be shown
that on-shell ordinary passive diffeomorphisms and the on-shell
Legendre pull-back of Hamiltonian gauge transformations are two
(overlapping) dense subsets of this set of on-shell
metric-dependent coordinate transformations. Since the Cauchy
surface for the Hole Argument lies outside the hole (where the
active diffeomorphism is the identity), it follows that the
passive re-interpretation of the active diffeomorphism $D^*_A$
must be {\it an on-shell Hamiltonian gauge transformation}, so
that Leibniz equivalence is identified with gauge equivalence in
the sense of Dirac constraint theory (${}^4g$ and $D^*_A\, {}^4g$
belong to the same gauge orbit).

\bigskip

What remains to be done is to implement Stachel's suggestion
according to which  the {\it intrinsic pseudo-coordinates} of
Bergmann and Komar \cite{54a} should be used as individuating
fields. These pseudo-coordinates for $M^4$ (at least when there
are no Killing vectors) are four scalar functions
$F^A[w_{\lambda}]$, $A, \lambda = 1,..,4$, of the four eigenvalues
$w_{\lambda}({}^4g, \partial\, {}^4g)$ of the spatial part of the
Weyl tensor. Since these eigenvalues can be shown to be in general
functions of the 3-metric, of its conjugate canonical momentum
(namely of the extrinsic curvature of $\Sigma_{\tau}$) and of the
lapse and shift functions, the pseudo-coordinates are well defined
in phase space and can be used as a label for the points of $M^4$.

\bigskip

The final step \cite{55a} is to implement the individuation of
point-events by considering an arbitrary admissible 3+1 splitting
of $M^4$ with a given time-like observer and the associated radar
4-coordinates $\sigma^A$ and by imposing the following gauge
fixings to the secondary super-hamiltonian and super-momentum
constraints (the only restriction on the functions $F^A$ is the
orbit condition)

\beq
 \chi^A(\tau ,\sigma^r) = \sigma^A - F^A[w_{\lambda}] \approx 0.
 \nonumber \\
 \eeq

In this way we break completely general covariance and we
determine the gauge variables $\xi^r$ and $\pi_{\phi}$. Then the
$\tau$-constancy of these gauge fixings will produce the gauge
fixings determining the lapse and shift functions. After having
fixed the Lorentz gauge freedom of the cotetrads, we arrive at a
completely fixed gauge in which, after the transition to Dirac
brackets, we get $\sigma^A \equiv {\tilde F}^A[r_{\bar a}(\sigma
), \pi_{\bar a}(\sigma )]$, namely that the radar 4-coordinates of
a point in $M^4_{3+1}$, the copy of $M^4$ coordinatized with the
chosen non-inertial frame, are determined {\it off-shell} by the
four DO of that gauge: in other words {\it the individuating
fields are the genuine tidal effects of the gravitational field}.
By varying the functions $F^A$ we can make an analogous off-shell
identification in every other admissible non-inertial frame. The
procedure is consistent, because the DO know the whole 3+1
splitting $M^4_{3+1}$ of $M^4$, being functionals not only of the
3-metric on $\Sigma_{\tau}$, but also of its extrinsic curvature.

\bigskip

Some consequences of this identification of the point-events of
$M^4$ are:\medskip

1) The space-time $M^4$ and the gravitational field are
essentially the same entity. The presence of matter modifies the
solutions of Einstein equations, i.e. $M^4$, but does not play any
role in this identification. Instead matter is fundamental for
establishing a (still lacking) dynamical theory of measurement not
using test objects. As a consequence, instead of the dichotomy
substantivalism/relationism, it seems that this analysis - as a
case study limited to the class of space-times dealt with - may
offer a new more articulated point of view, which can be named
{\it point structuralism} (see Ref. \cite{56a}).

\medskip

2) The reduced phase space of this model of general relativity is
the space of abstract DO (pure tidal effects without inertial
effects), which can be thought as four fields on an abstract
space-time ${\tilde M}^4 = \{ equivalence\, class\, of\, all\,
the\, admissible\, non-inertial\, frames\, M^4_{3+1}\,
containing\, the\, associated\, inertial\, effects\}$.

\medskip

3) Each radar 4-coordinate system of an admissible non-inertial
frame $M^4_{3+1}$ has an associated {\it non-commutative
structure}, determined by the Dirac brackets of the functions $
{\tilde F}^A[r_{\bar a}(\sigma ), \pi_{\bar a}(\sigma )]$
determining the gauge.

\medskip

4) Conjecture: there should exist privileged Shanmugadhasan
canonical bases of phase space, in which the DO (the tidal
effects) are also {\it Bergmann observables} \cite{57a}, namely
coordinate-independent (scalar) tidal effects.

\bigskip

As a final remark, let us note that these results on the
identification of point-events are {\it model dependent}. In
spatially compact space-times without boundary, the DO are {\it
constants of the motion} due to the frozen picture. As a
consequence, the gauge fixings $\chi^A(\tau ,\sigma^r) \approx 0$
(in particular $\chi^{\tau}$) cannot be used to rebuild the
temporal dimension: probably only the instantaneous 3-space of a
3+1 splitting can be individuated in this way.

\section{Future Developments.}

I will finish with a list of the open problems in canonical metric
and tetrad gravity for which there is a concrete hope to be
clarified and solved in the near future.
\bigskip

i) A different Shanmugadhasan canonical transformation, adapted
only to 10 constraints but allowing the addition of any kind of
matter to the rest-frame instant form of tetrad gravity, has been
recently found starting from a new parametrization of the 3-metric
\cite{58a}. This transformation is the first explicit construction
of a York map \cite{59a}, in which the momentum conjugate to the
conformal factor (the gauge variable controlling the convention
for clock synchronization) is proportional to the trace
${}^3K(\tau ,\vec \sigma )$ of the extrinsic curvature of the
simultaneity surfaces $\Sigma_{\tau}$. Both the tidal variables
and the gauge ones can be expressed in terms of the original
variables. The solution of the super-momentum constraints shows
the existence of a generalized Gribov ambiguity connected with the
gauge group of 3-diffeomorphisms. Also the Hamiltonian
interpretation of harmonic gauges is given. Moreover, in a family
of completely fixed gauges differing for the convention of clock
synchronization, the deterministic Hamilton equations for the
tidal variables and for the matter contain relativistic inertial
forces determined by ${}^3K(\tau ,\vec \sigma )$, which change
from attractive to repulsive where the trace change sign. These
inertial forces do not have a non-relativistic counterpart (the
Newton 3-space is absolute) and could support the proposal of Ref.
\cite{60a} \footnote{The model proposed in Ref.\cite{60a} is too
naive as shown by the criticism in Refs.\cite{61a}.} that {\it
dark matter} can be explained as an {\it inertial effect}. Also it
would be interesting to have some understanding of how is
distributed the gravitational energy density in different
coordinate systems and how it depends on the convention for clock
synchronization. Has this distribution any relevance for the dark
energy problem?

\bigskip

ii) The York basis will allow to study the weak-field
approximation to the two-body problem in a post-Minkowskian
background-independent way by using a Grassmann regularization of
the self-energies, following the track of Refs. \cite{25a}. The
solution of the Lichnerowicz equation would allow to find the
expression of {\it the relativistic Newton and gravito-magnetic
action-at-a-distance potentials} between the two bodies (sources,
among other effects, of the {\it Newtonian tidal effects}) and the
coupling of the particles to the DO of the gravitational field
(the genuine tidal effects) in various radar coordinate systems:
it would amount to a re-summation of the $1/c$ expansions of the
Post-Newtonian approximation. Also the relativistic version of the
quadrupole formula for the emission of gravitational waves from
the binary system could be obtained.  Moreover, we could
understand better what replaces the spin-2 approximation of
gravity on a fixed background space-time in a background-
independent scheme and in non-harmonic gauges. Finally one could
try to define a relativistic gravitational micro-canonical
ensemble generalizing the Newtonian one developed in
Ref.\cite{62a}.
\bigskip

iii) With more general types of matter (relativistic fluids
\cite{23a}, scalar \cite{21a} and electromagnetic \cite{12a}
fields) it should be possible to develop Hamiltonian numerical
gravity based on the Shanmugadhasan canonical bases and to study
post-Minkowskian approximations based on power expansions in
Newton constant. Moreover one should look for strong-field
approximations to be used in the gravitational collapse of a ball
of fluid.

\bigskip

iv) Find the Hamiltonian formulation of the Newman-Penrose
formalism (see Ref.\cite{63a}), in particular of the 10 Weyl
scalars. Look for the Bergmann observables (the scalar tidal
effects) and try to understand {\it which inertial effects may
have a coordinate-independent form and which are  intrinsically
coordinate-dependent like the ADM energy density}. Look for the
existence of a closed Poisson algebra of scalars and for
Shanmugadhasan canonical bases incorporating the Bergmann
observables, to be used to find new expressions for the
super-hamiltonian and super-momentum constraints, hopefully easier
to be solved.

\bigskip

v) Find the clock synchronization convention hidden in the
Post-Newtonian metric used around the geoid for space navigation
and for GPS: it deviates from Einstein's convention at the order
$1/c^2$. Moreover try to understand which notions of instantaneous
3-space and of simultaneity are implied by the luminosity distance
used in astrophysics and cosmology.
\bigskip

vi) Find all the admissible 3+1 splittings of Minkowski space-time
which avoid the Torre-Varadarajan no-go theorem. Then adapt these
3+1 splittings to tetrad gravity and try to see whether it is
possible to arrive at a multi-temporal background- and coordinate-
independent quantization of the gravitational field, in which only
the Bergmann observables (the scalar tidal effects) are quantized,
maybe by using the non-commutative structure associated to each
non-inertial system.
\bigskip

vii) Find the special relativistic version of Bell inequalities
and the role of the  non-local notions of clock synchronization
convention and of separation of the relativistic center of mass
from relative motions in the problem of entanglement relying on
Ref.\cite{20a}.

\appendix

\section{Dirac's Constraint Theory.}

Dirac's theory of constraints \cite{a1,a2} is needed for the
Hamiltonian formulation of special relativistic systems, gauge
theories and general relativity, whose configuration description
requires the use of singular Lagrangians with a degenerate Hessian
matrix. The second Noether theorem is the basic tool to understand
the properties of the associated Euler-Lagrange equations
\cite{a3}.\medskip

The local invariances (or quasi-invariances) of the singular
Lagrangians under local Noether transformations, depending upon
arbitrary functions of time or space-time, imply that there are
certain arbitrary velocities. The Noether identities propagate
this indetermination to other variables in the case of first-class
constraints, while they lead to a final elimination of the
arbitrariness in the case of the second-class ones. In the case of
first class constraints the local Noether transformations are
called gauge transformations. As a consequence, when first and
second class constraints are present, some variables, the {\it
gauge variables}, remain arbitrary; others are completely
determined and can be eliminated; finally the true physical
degrees of freedom, satisfying deterministic equations of motion,
are the gauge invariant quantities, the {\it Dirac observables}
(DO). One of the basic problems is the separation of these three
types of quantities starting from the initial configuration
variables. The only known algorithm for realizing this separation
are the {\it Shanmugadhasan canonical transformations}
\cite{a4,a5}. These transformations are implicitly assumed in the
definition of the Faddeev-Popov measure \cite{a6} of the path
integral and in the BRST method \cite{a7}.

\bigskip

Let us review some of the main properties of constrained systems.
\medskip

If a finite-dimensional system with configuration space $Q$
\footnote{$q^i$, i=1,..,N, are local coordinates in a global
(assumed to exist for the sake of simplicity) chart of the atlas
of Q; $(t,q^i(t))$ is a point in $R\times Q$, where $R$ is the
time axis; ${\dot q}^i(t)=dq^i(t)/dt$} is described by a singular
Lagrangian $L$, namely with a degenerate Hessian matrix ($det\,
\Big( \partial^2L/\partial {\dot q}^i\partial {\dot q}^j\Big)
=0$), its Euler-Lagrange (EL) equations are in general a mixture
of three types of equations

\noindent i) some depending only on the $q^i$ (holonomic
constraints);

\noindent ii) some depending only on $q^i$ and ${\dot q}^i$
(Lagrangian, in general non-holonomic, constraints and/or, when
non projectable to phase space, intrinsic first order equations of
motion violating the so called second order differential equation
(SODE) conditions);

\noindent iii) some depending on $q^i$, ${\dot q}^i$, ${\ddot
q}^i$ (genuine second order equations of motion, which however
cannot be put in normal form, i.e. solved in the ${\ddot q}^i$).

\noindent The solutions of the EL equations depend on arbitrary
functions of time, namely they are not deterministic.

\medskip

As shown in Refs.\cite{a3,a5}, to each null eigenvalue of the
Hessian matrix is associated a (quasi-) invariance of the action
principle under a set of local Noether transformations depending
upon an arbitrary function of time and a certain number of its
time derivatives. As a consequence to each null eigenvalue is
associated a chain of Noether identities, each one being the time
derivative of the previous one, and a undetermined primary
generalized velocity function. The Noether identities identify
more equations of the types i) and ii), as combinations of the
Euler-Lagrange equations and their time derivatives, and identify
secondary undetermined velocity (or configuration) functions. In
the simplest cases the chains are of two types:

\noindent i) the first class ones, in which the primary and
secondary velocity functions remain undetermined and the last
identity is a combination of the EL equations and their time
derivatives (contracted Bianchi identities, reducing the number of
independent equations of motion);

\noindent ii) the second-class ones, in which the last identity
determines the originally arbitrary primary and secondary velocity
functions.

Ordinary gauge theories are of the first type.

\medskip

The canonical momenta $p_i=\partial L/\partial {\dot q}^i$ are not
independent: there are relations among them
$\phi_{\alpha}(q,p)\approx 0$, called {\it primary Hamiltonian
constraints}, which define a sub-manifold $\gamma$ of the
cotangent space $T^{*}Q$ \footnote{The model is defined only on
this sub-manifold; one uses the Poisson brackets of $T^{*}Q$ in a
neighborhood of $\gamma$ and Dirac's weak equality $\approx$ means
that the equality sign cannot be used inside Poisson brackets}.
The canonical Hamiltonian $H_c(q,p)$ has to be replaced by the
Dirac Hamiltonian $H_D=H_c+\sum_{\alpha}
\lambda_{\alpha}(t)\phi_{\alpha}$, which knows the restriction to
the sub-manifold $\gamma$. The arbitrary Dirac multipliers
$\lambda_{\alpha}(t)$ are the phase space analogue of the
arbitrary primary velocity functions \footnote{Since there is no
canonical form of these velocity functions, it is convenient to
identify them with those functions which are mapped into the Dirac
multipliers by the Legendre transform. This prescription
eliminates every ambiguity in the Legendre transform \cite{a5}}.
The  consistency requirement that  the primary constraints are
preserved in time, $\partial_t \phi_{\alpha}=\{
\phi_{\alpha},H_D\} \approx 0$, either produces {\it secondary
Hamiltonian constraints} or determines some of the Dirac
multipliers. This procedure is repeated for the secondary
constraints \footnote{This is the {\it Dirac- Bergmann algorithm},
which can be shown to correspond to the projection to phase space
of the Noether identities \cite{a3,a5}.} and so on. At the end
there is a final set of constraints $\chi_a\approx 0$ defining the
final sub-manifold $\bar \gamma$ of $T^{*}Q$ on which the dynamics
is consistently restricted, and a final Dirac Hamiltonian with a
reduced set of arbitrary Dirac multipliers describing the
remaining indetermination of the time evolution. The constraints
are divided into two subgroups:

\noindent i) the {\it first class} ones $\chi^{(1)}_m\approx 0$,
having weakly zero Poisson bracket with all constraints and being
the generators of the Hamiltonian version of the gauge
transformations of the theory (the associated vector fields $\{ .,
\chi^{(1)} _m\}$ are tangent to $\bar \gamma$);

\noindent ii) the {\it second class} ones $\chi^{(2)}_n \approx 0$
(their number is even) with $det\, \Big( \{ \chi^{(2)}_{n_1},
\chi^{(2)}_{n_2}\} \Big) \not= 0$, corresponding to pairs of
inessential eliminable variables (the associated vector fields are
normal to $\bar \gamma$).

\medskip

The solutions of the Hamilton-Dirac equations with the final Dirac
Hamiltonian depend on as many arbitrary functions of time as the
final undetermined Dirac multipliers. The vector fields associated
with the first class constraints generate a foliation of the
sub-manifold $\bar \gamma$: each leaf ({\it Hamiltonian gauge
orbit}) contains all the configurations which are gauge equivalent
and which have to be considered as the same physical configuration
({\it equivalence class of gauge equivalent configurations}); the
canonical Hamiltonian $H_c$ (if it is not $H_c\approx 0$)
generates an evolution which maps one gauge orbit into the others.

\medskip

 Therefore, the physical reduced phase space is obtained: i)
by eliminating as many pairs of conjugate variables as second
class constraints by means of the so called associated Dirac
brackets; ii) by going to the quotient with respect to the
foliation (a representative of the reduced phase space can be
build by adding as many gauge-fixing constraints as first class
ones, so to obtain a set of second class constraints if an orbit
condition is satisfied). In general this procedure breaks the
original manifest Lorentz covariance in the case of special
relativistic systems.

\medskip

Let us remark that only the primary first class constraints are
associated with arbitrary Dirac multipliers.

\medskip

The natural way to add gauge-fixing constraints when there are
secondary first class constraints, is to start giving the gauge
fixings to the secondary constraints. The consistency requirement
that the gauge fixings be preserved in time will generate the
gauge fixings for the primary first class constraints and the time
constancy of these new gauge fixings will determine the Dirac
multipliers eliminating every residual gauge freedom. The same
method holds with chains of first class constraints of any length.

\medskip

The DO are the {\it gauge invariant} functions on the reduced
phase space, on which there is a deterministic evolution generated
by the projection of the canonical Hamiltonian. Therefore, the
main problem is to find a (possibly global) Darboux coordinate
chart of the reduced phase space, namely a canonical basis of DO.

\medskip

When there is reparametrization invariance of the original action,
the canonical Hamiltonian vanishes and the reduced phase space is
said frozen (like it happens in Hamilton-Jacobi theory). When this
happens,  both kinematics and dynamics are contained in the first
class constraints describing the system: these can be interpreted
as generalized Hamilton-Jacobi equations \cite{a8}, so that the DO
turn out to be the Jacobi data. When there is a kinematical
symmetry group, like the Galileo or Poincar\'e groups, an
evolution may be reintroduced by using the energy generator as
Hamiltonian \cite{a3}.

\medskip

\bigskip

Now I will delineate the main steps for the determination of the
DO in the case in which only {\it primary first class} constraints
$\phi _{\alpha}\approx 0$ are present at the Hamiltonian level.

\medskip

The Euler-Lagrange equations associated with a singular Lagrangian
do not determine the gauge part of the extremals. However it
cannot be totally arbitrary, but must be compatible with the
algebraic properties of the Noether gauge transformations induced
by the first class constraints under which the action is either
invariant or quasi-invariant as implied by the second Noether
theorem. In the Hamiltonian formulation these properties are
contained in the structure constants, or functions, of the Poisson
brackets of the first-class constraints among themselves [$\{
\phi_{\alpha}, \phi _{\beta}\} = C_{\alpha\beta\gamma}
\phi_{\gamma}$, $\{ \phi_{\alpha},H_c\} =
C_{\alpha\beta}\phi_{\beta}$] and the gauge arbitrariness of the
trajectories is described by the Dirac multipliers appearing in
the Dirac Hamiltonian. In both formulations one has to add extra
equations, the either Lagrangian or Hamiltonian {\it
multi-temporal equations} \cite{a9}, to have a consistent
determination of the gauge part of the trajectory. These equations
are obtained by rewriting the variables $q^i(t)$, $p_i(t)$ in the
form $q^i(t,\tau_{\alpha})$, $p_i(t,\tau_{\alpha})$, and by
assuming that the original t-evolution generated by the Dirac
Hamiltonian $H_D=H_c+\sum_{\alpha}\lambda_{\alpha}(t)
\phi_{\alpha}$ is replaced by:

\noindent i) a deterministic t-evolution generated by $H_c$;

\noindent ii) a $\tau_{\alpha}$-evolution (re-assorbing the
arbitrary Dirac multipliers $\lambda_{\alpha}(t)$), for each
$\alpha$, generated in a suitable way by the first class
constraints $\phi_{\alpha}$.
\medskip

The $\tau_{\alpha}$-dependence of $q^i$, $p_i$ determined by these
multi-temporal (or better multi-parametric) equations, which are
integrable due to the first-class property of the constraints,
describes their dependence on the gauge orbit, once  Cauchy data
for the Hamilton-Dirac equations for the DO are given.

\medskip

When the Poisson brackets of the Hamiltonian first class
constraints imply a canonical realization of a Lie algebra, the
extra Hamiltonian multi-temporal equations have the first class
constraints as Hamiltonians and the time parameters (replacing the
Dirac multipliers) are the coordinates of a group manifold for a
Lie group whose algebra is the given Lie algebra: they enter in
the multi-temporal equations via a set of left invariant vector
fields $Y_{\alpha}$ on the group manifold [$Y_{\alpha} A(q,p)=\{
A(q,p), \phi_{\alpha}\}$]. In the ideal case in which the gauge
foliation of $\bar \gamma$ is nice, all the leaves (or gauge
orbits) are diffeomorphic and, in the simplest case, all of them
are diffeomorphic to the group manifold of a Lie group. In this
ideal case to rebuild a gauge orbit from one of its points (and
therefore to determine the gauge part of the trajectories passing
through that point) one needs the Lie equations associated with
the given Lie group: {\it the Hamiltonian multi-temporal equations
are generalized Lie equations describing all the gauge orbits
simultaneously}. In a generic case this description holds only
locally for a set of diffeomorphic orbits, also in the case of
systems invariant under diffeomorphisms.

\medskip

Once one has solved the multi-temporal equations, the next step is
the determination of a  {\it Shanmugadhasan canonical
transformation} \cite{a4,a5}. In the finite dimensional case
general theorems \cite{a10} connected with the Lie theory of
function groups \cite{a11} ensure the existence of local canonical
transformations from the original canonical variables $q^i$,
$p_i$, in terms of which the first class constraints (assumed
globally defined) have the form $\phi_{\alpha}(q,p)\approx 0$, to
canonical bases $P_{\alpha}$, $Q_{\alpha}$, $P_A$, $Q_A$, such
that

i) the equations $P_{\alpha}\approx 0$ locally define the same
original constraint manifold (the $P_{\alpha}$ are an
Abelianization of the first class constraints);

ii) the $Q_{\alpha}$ are the adapted Abelian gauge variables
describing the gauge orbits (they are a realization of the times
$\tau_{\alpha}$ of the multi-temporal equations in terms of
variables $q^i$, $p_i$);

iii) the $Q_A$, $P_A$ are an adapted canonical basis of DO.

\medskip

Therefore the problem of the search of the DO becomes the problem
of finding Shanmugadhasan canonical transformations. The strategy
is to find Abelianizations $P_{\alpha}$ of the original
constraints, to solve the multi-temporal equations for $q^i$,
$p_i$ associated with the $P_{\alpha}$, to determine the
multi-times $Q_{\alpha}=\tau_{\alpha}$ and to identify the DO
$P_A$, $Q_A$ from the remaining original variables, i.e. from
those their combinations independent from $P_{\alpha}$ and
$Q_{\alpha}$.

\bigskip

In gauge field theories the situation is more complicated, because
the theorems ensuring the existence of the Shanmugadhasan
canonical transformation have not been extended to the
infinite-dimensional case and one must use heuristic
extrapolations of them. One of the reasons is that some of the
constraints can now be interpreted as elliptic equations and they
can have zero modes. Let us consider the stratum $\sgn\, p^2 > 0$
of free Yang-Mills theory as a prototype and its first class
constraints, given by the Gauss laws and by the vanishing of the
time components of the canonical momenta. The problem of the {\it
zero modes} will appear as a {\it singularity structure of the
gauge foliation} of the allowed strata, in particular of the
stratum $\sgn\, p^2 > 0$. This phenomenon was discovered in
Ref.\cite{a12} by studying the space of solutions of Yang-Mills
and Einstein equations, which can be mapped onto the constraint
manifold of these theories in their Hamiltonian description. It
turns out that the space of solutions has a {\it cone over cone}
structure of singularities: if we have a line of solutions with a
certain number of symmetries, in each point of this line there is
a cone of solutions with one less symmetry. In the Yang-Mills case
the {\it  gauge symmetries} of a gauge potential are connected
with the generators of its stability group, i.e. with the subgroup
of those special gauge transformations which leave invariant that
gauge potential (this is the Gribov ambiguity for gauge
potentials; there is also a more general Gribov ambiguity for
field strengths, the {\it gauge copies} problem; for all these
problems see Ref. \cite{27a} and its bibliography).

Since the Gauss laws are the generators of the gauge
transformations (and depend on the chosen gauge potential through
the covariant derivative), this means that for a gauge potential
with non trivial stability group those combinations of the Gauss
laws corresponding to the generators of the stability group cannot
be any more first class constraints, since they do not generate
effective gauge transformations but special symmetry
transformations. This problematic has still to be clarified, but
it seems that in this case these components of the Gauss laws
become {\it third class constraints}, which are not generators of
true gauge transformations. This new kind of constraints was
introduced in Refs.\cite{a3,a5} in the finite dimensional case as
a result of the study of some examples, in which the Jacobi
equations (the linearization of the Euler-Lagrange equations) are
singular, i.e. some of their solutions are not infinitesimal
deviations between two neighboring extremals of the Euler-Lagrange
equations. This interpretation seems to be confirmed by the fact
that the singularity structure discovered in Ref.\cite{a12}
follows from the existence of singularities of the linearized
Yang-Mills and Einstein equations. These problems are part of the
Gribov ambiguity, which, as a consequence, induces an extremely
complicated stratification and also singularities in each
Poincar\'e stratum of $\bar \gamma$.

\medskip

Other possible sources of singularities of the gauge foliation of
Yang-Mills theory in the stratum $\sgn\, p^2 > 0$ may be:

\noindent i) different classes of gauge potentials identified by
different values of the field invariants;

\noindent ii) the orbit structure of the rest frame (or Thomas)
spin $\vec S$, identified by the Pauli-Lubanski Casimir $W^2 = -
\sgn\, p^2\, {\vec S}^2$ of the Poincare' group.

\medskip

The final outcome of this structure of singularities is that the
reduced phase-space, i.e. the space of the gauge orbits, is in
general a {\it stratified manifold with singularities} \cite{a12}.
In the stratum $\sgn\, p^2 > 0$ of the Yang-Mills theory these
singularities survive the Wick rotation to the Euclidean
formulation and it is not clear how the ordinary path integral
approach and the associated BRS method can take them into account
(they are zero measure effects). The search of a global canonical
basis of DO for each stratum of the space of the gauge orbits can
give a definition of the measure of the phase space path integral,
but at the price of a non polynomial Hamiltonian. Therefore, if it
is not possible to eliminate the Gribov ambiguity (assuming that
it is only a mathematical obstruction without any hidden physics),
the existence of global DO for Yang-Mills theory is very
problematic.

\bigskip

See Ref.\cite{58a} for the recently discovered generalized Gribov
ambiguity in metric and tetrad gravity arising in the solution of
the super-momentum constraints after having done the York map.

\vfill\eject

\end{document}